\documentclass{amsart}

\usepackage{amsmath, amsthm, amssymb, amsfonts, graphicx, slashed, mathabx, url}
\usepackage{latexsym}

\newtheorem{notation}{Notation}
\newtheorem*{hypothesis}{Hypothesis}

\newcommand{\Z}{\ensuremath{\mathbb{Z}}}
\newcommand{\R}{\ensuremath{\mathbb{R}}}
\newcommand{\C}{\ensuremath{\mathbb{C}}}

\newcommand{\CI}{\ensuremath{\mathbb{C}_I}}
\newcommand{\Hh}{\ensuremath{\mathbb{H}}}

\newcommand{\CIo}{\ensuremath{\mathbb{C}_I\backslash\{0\}}}		
\newcommand{\CIod}{\ensuremath{\mathbb{C}_I^2\backslash\{0\}}}	
\newcommand{\CIof}{\ensuremath{\mathbb{C}_I^4\backslash\{0\}}}		
\newcommand{\Ho}{\ensuremath{\mathbb{H}\backslash\{0\}}}			
\newcommand{\Hod}{\ensuremath{\mathbb{H}^2\backslash\{0\}}}		

\newcommand{\CP}{\ensuremath{\mathbb{CP}}}
\newcommand{\HP}{\ensuremath{\mathbb{HP}}}
\newcommand{\CPI}{\ensuremath{\mathbb{CP}_I}}

\newcommand{\I}{\ensuremath{\bar{I}}}
\newcommand{\J}{\ensuremath{\bar{J}}}
\newcommand{\K}{\ensuremath{\bar{K}}}
\newcommand{\Ip}{\ensuremath{\bar{I}^+}}
\newcommand{\Jp}{\ensuremath{\bar{J}^+}}
\newcommand{\Kp}{\ensuremath{\bar{K}^+}}

\newcommand{\HL}{\ensuremath{\mathcal{H}_\lambda}}				
\newcommand{\ML}{\ensuremath{\mathcal{M}_\lambda}}			
\newcommand{\MLc}{\ensuremath{\mathcal{M}_\lambda^c}}		
\newcommand{\TLI}{\ensuremath{\mathcal{T}_{\lambda I}}}			

\newcommand{\Th}{\ensuremath{T_I^{1,0}}}					
\newcommand{\Tah}{\ensuremath{T_I^{0,1}}}					

\newcommand{\Zu}{\ensuremath{Z^\mu}}					
\newcommand{\Zubar}{\ensuremath{\bar{Z}^\mu}}					
\newcommand{\Tu}{\ensuremath{T^\mu}}					
					
\newcommand{\Vu}{\ensuremath{V^\mu}}					
					
\newcommand{\Zo}{\ensuremath{Z^0}}					
\newcommand{\Zobar}{\ensuremath{\bar{Z}^0}}					
\newcommand{\To}{\ensuremath{T^0}}					
\newcommand{\Tobar}{\ensuremath{\bar{T}^0}}					
\newcommand{\Vo}{\ensuremath{V^0}}					
\newcommand{\Vobar}{\ensuremath{\bar{V}^0}}					

\newcommand{\Qu}{\ensuremath{Q_n^\mu}}					
\newcommand{\Qubar}{\ensuremath{\bar{Q}_n^\mu}}					
\newcommand{\Du}{\ensuremath{D_n^\mu}}					
\newcommand{\Dubar}{\ensuremath{\bar{D}_n^\mu}}					
\newcommand{\Su}{\ensuremath{S_n^\mu}}					
\newcommand{\Subar}{\ensuremath{\bar{S}_n^\mu}}					
\newcommand{\Qo}{\ensuremath{Q_n^0}}					
\newcommand{\Qobar}{\ensuremath{\bar{Q}_n^0}}					
\newcommand{\Do}{\ensuremath{D_n^0}}					
\newcommand{\Dobar}{\ensuremath{\bar{D}_n^0}}					
\newcommand{\So}{\ensuremath{S_n^0}}					
\newcommand{\Sobar}{\ensuremath{\bar{S}_n^0}}					
\newcommand{\Dio}{\ensuremath{D_I^0}}					
\newcommand{\Sio}{\ensuremath{S_I^0}}					

\newcommand{\Ps}{\ensuremath{\psi_g}}					
\newcommand{\Pst}{\ensuremath{\widetilde{\psi_g}}}					
\newcommand{\PsL}{\ensuremath{\psi_g^L}}					
\newcommand{\PsR}{\ensuremath{\psi_g^R}}					

\newcommand{\PsTL}{\ensuremath{\bar{\psi}_g^L}}			
\newcommand{\PL}{\ensuremath{\psi_{g \lambda}^-}}		
\newcommand{\PLp}{\ensuremath{\psi_{g \lambda}^+}}		

\newcommand{\wk}[2]{\ensuremath{\nabla_{b #1}^w {#2}}}				
\newcommand{\wkc}[2]{\ensuremath{\overline{\nabla}_{b #1}^w {#2}}}	
\newcommand{\whk}[2]{\ensuremath{\nabla_{h #1}^w {#2}}}				
\newcommand{\lk}[2]{\ensuremath{\nabla_{b #1}^{LC} {#2}}}			
\newcommand{\lkc}[2]{\ensuremath{\overline{\nabla}_{b #1}^{LC} {#2}}}			
\newcommand{\lhk}[2]{\ensuremath{\nabla_{h #1}^{LC} {#2}}}			
\newcommand{\spl}[2]{\ensuremath{\nabla_{g #1}^{sp} {#2}}}			
\newcommand{\lcl}[2]{\ensuremath{\nabla_{g #1}^{LC} {#2}}}			
\newcommand{\wksl}[2]{\ensuremath{\nabla_{bg #1}^{ws} {#2}}}		
\newcommand{\wkslc}[2]{\ensuremath{\overline{\nabla}_{bg #1}^{ws} {#2}}}
\newcommand{\whksl}[2]{\ensuremath{\nabla_{hg #1}^{ws} {#2}}}		
\newcommand{\lfs}[2]{\ensuremath{\nabla_{k #1}^{LC} {#2}}}				
\newcommand{\lfsc}[2]{\ensuremath{\overline{\nabla}_{k #1}^{LC} {#2}}}	
\newcommand{\lcq}[2]{\ensuremath{\nabla_{q #1}^{LC} {#2}}}				

\newcommand{\dtc}{\ensuremath{\dot{\theta}}}		
\newcommand{\dth}{\ensuremath{\dot{\theta}}}		
\newcommand{\dtcl}{\ensuremath{\dot{\theta}^L}}		
\newcommand{\dthl}{\ensuremath{\dot{\theta}^L}}		
\newcommand{\dtcr}{\ensuremath{\dot{\theta}^R}}		
\newcommand{\dthr}{\ensuremath{\dot{\theta}^R}}		
\newcommand{\Higgs}{\ensuremath{\theta_{\Zo} + \theta_{\Qo}}}		

\newcommand{\Db}{\ensuremath{\slashed{\nabla}_{k}}}		
\newcommand{\Dbc}{\ensuremath{\overline{\slashed{\nabla}}_{k}}}		
\newcommand{\D}{\ensuremath{\slashed{\nabla}_1}}					
\newcommand{\Dc}{\ensuremath{\overline{\slashed{\nabla}}_1}}		
\newcommand{\Dh}{\ensuremath{\slashed{\nabla}_q}}				
\newcommand{\dd}{\ensuremath{\slashed{\nabla}_2}}					
\newcommand{\DbhL}{\ensuremath{\slashed{\nabla}_{kq}^L}}		
\newcommand{\DbR}{\ensuremath{\slashed{\nabla}_{k2}^R}}		
\newcommand{\DbLc}{\ensuremath{\overline{\slashed{\nabla}}_{k2}^L}}		
\newcommand{\DbhRc}{\ensuremath{\overline{\slashed{\nabla}}_{kq}^R}}		
\newcommand{\DhL}{\ensuremath{\slashed{\nabla}_{1q}^L}}				
\newcommand{\DR}{\ensuremath{\slashed{\nabla}_{12}^R}}					
\newcommand{\DLc}{\ensuremath{\bar{\slashed{\nabla}}_{12}^L}}		
\newcommand{\DhRc}{\ensuremath{\overline{\slashed{\nabla}}_{1q}^R}}		

\newcommand{\gam}{\ensuremath{\gamma^\mu}}		
\newcommand{\sig}{\ensuremath{\sigma^\mu}}			
\newcommand{\sigr}{\ensuremath{\sigma^{\mu *}}}		
\newcommand{\xc}{\ensuremath{\otimes_{\C_i}}}		
\newcommand{\xj}{\ensuremath{\otimes_J}}				
\newcommand{\half}{\tfrac{1}{2}}

\thanks{The author thanks Alexia, Allison, Morgan, Joanne and Joan for their encouragement and 
support.}

\begin{document}

\title{A Moduli Space of the Quaternionic Hopf Surface Encodes Standard Model Physics}
\author{Colin B. Hunter}
\email{colinh@alumni.stanford.edu}
\date{September 13, 2012}

\begin{abstract}
The quaternionic Hopf surface, \HL, is associated with a non-compact moduli space, \ML, 
of stable holomorphic $SL(2,C)$ bundles. $\ML$ is open in \MLc, the corresponding compact 
moduli space of holomorphic $SL(2,C)$ bundles, and naturally fibers over an open set of 
the quaternionic projective line $\HP^1$. We pull back to $\ML$ natural locally conformal 
kaehler and hyperkaehler structures from \MLc, and lift natural sub-pseudoriemannian 
and optical structures from $\HP^1$. Unexpectedly, the holomorphic maps connecting these 
structures solve the the classical Dirac-Higgs equations of the unbroken Standard 
Model. These equations include: all observed fermionic and bosonic fields of all three generations 
with the correct color, weak isospin, and hypercharge values; a Higgs field coupling left 
and right fermion fields; and a pp-wave gravitational metric. We hypothesize that physics is 
essentially the geometry of \ML, both algebraic (quantum) and differential (classical). We 
further show that the Yang-Mills equations with fermionic currents also naturally emerge, 
along with an induced action on the $\ML$ structure sheaf equivalent to the time-evolution operator 
of the associated quantum field theory. 
\end{abstract}

\maketitle

\setcounter{tocdepth}{2}
\tableofcontents

\section{Introduction}\label{S:intro} 

Braam and Hurtubise constructed the moduli spaces of stable holomorphic $SL(2,C)$ bundles on elliptic Hopf surfaces \cite{pb89}. These moduli spaces are indexed by the second Chern class of the bundles $c2$, ($c1 = 0$, as the bundles have trivial determinant, since the structure group is $SL(2,C)$.) The simplest of these moduli spaces has $c2 =1$ (the $c2 = 0$ case is trivial). 

For the ``hyper-elliptic'' quaternionic Hopf surface, $\HL$, defined by
\[
\Ho/\lambda^n ; \lambda \colon q \to \lambda q ;  \lambda \in \R > 1, q \in \Hh
\]
we shall call the associated ($c2 = 1$) moduli space \ML. 

Both $\HL$ and $\ML$ are quite interesting mathematically. Firstly, $\HL$ is non-Kaehlerian and non-algebraic (as are all Hopf surfaces); in fact Hopf surfaces are the standard example of a non-Kaehlerian, non-algebraic manifold. Secondly, quaternionic Hopf surfaces are hypercomplex manifolds. (In fact, quaternionic Hopf surfaces are among the very few such hypercomplex four-manifolds \cite{mk80}. Thirdly, $\HL$ is also a group manifold, inheriting the non-abelian algebraic structure of the quaternions, just as elliptic curves inherit the abelian group structure of the complex numbers. In fact, quaternionic Hopf surfaces are in many ways the best quaternionic equivalent of elliptic curves.  

$\ML$ is also extremely interesting mathematically. Like $\HL$, it is hypercomplex and hyperelliptic, but over an open set of $\CPI^3$. It also fibers naturally over an open set of $\HP^1$. It is usually called the space of ``instantons'' over \HL, since it is isomorphic to the moduli space of single instantons, i.e., anti-self-adjoint $SU(2)$ connections. The literature on instantons is vast, especially since mathematicians began to apply them to problems in geometry in the late 1970s with great success (notably, Atiyah-Hitchin-Singer \cite{ma78} and Donaldson \cite{sd90}). The theorem that expresses the isomorphism of stable $SL(2,C)$ holomorphic bundles with anti-self-adjoint $SU(2)$ connections evolved during the 1980s \cite{ku86}, but in its most general form it is often called the ``Kobayashi-Hitchin Correspondance'' \cite{ml95}. 

Verbitsky showed \cite{mv96} that, in the case of such bundles over hypercomplex four-manifolds, the anti-self-adjointness criterion is equivalent to Òhyperholomorphicity:Ó the bundles are holomorphic for all complex structures on the base. (Verbitsky actually used hyperkaehler surfaces, but Widdows \cite{dw00} showed that hypercomplexity is sufficient.) Thus, $\ML$ is also the lowest-order non-trivial moduli space of hyper-holomorphic \Hh-line bundles. As such, it is kind of the quaternionic equivalent of the lowest moduli space of holomorphic \C-line bundles over elliptic curves, that is, the Picard variety of the torus.  Thus, $\ML$ is a ``generalized Jacobian'' and should play a significant role in further explorations of quaternionic geometry. (We therefore employ ``hyperholomorphic \Hh-line bundle'' basically as a suggestive synonym for Òstable holomorphic $SL(2,C)$ bundleÓ.)

For the most part we will not discuss the mathematical interest in $\HL$ and $\ML$, except for a few speculations in the final section. Instead, we will explore an unexpected connection between the geometry of  $\ML$ and the fundamental equations of physics. Two natural sets of structures on $\ML$, one pulled back from a natural compactification, the other lifted from the base of a natural fibration, are related by harmonic maps. We find that the classical unbroken Dirac-Higgs equations of the Standard Model, for all fermions of all generations and all gauge connections, and all known interaction charges (color, weak isospin, and hypercharge), are naturally isomorphic to the tension equation of these harmonic maps. In addition, a naturally induced sub-pseudoriemannian metric solves the Einstein equations for a plausible (unbroken fields) stress-energy tensor. 

We therefore hypothesize that the Standard Model classical field theory is essentially the differential geometry of the moduli space \ML. We show that the associated Yang-Mills equations (with currents) emerge from the Weyl connectionÕs curvature properties, as does the equation for the Higgs field. We define a structure sheaf for $\ML$ that captures both its hypercomplex and elliptic structure and show that the coupled equations induce a holomorphic flow on $\ML$ and thus an action on this sheaf.  We propose that this action on the structure sheaf is the propagator on the state space of the associated quantum field theory, which is thus the corresponding algebraic geometry of \ML.

In section \ref{S:struct}, we will derive $\ML$, using the graph technique of Braam-Hurtubise, then define two sets of structures on \ML. First, we will define locally conformal kaehler (LCK) and locally conformal hyperkaehler (LCHK) structures on pulled back from \MLc, a compact moduli space in which $\ML$ is open. Then we will then define natural fibrations of $\ML$ over open sets of $\CPI^3$ and $\HP^1$, and a natural action by the $I$ automorphism group $SL(2,C_I)$. We will use this natural action to define several structures on the base manifold, in particular a Lorentzian metric that supports a non-zero parallel $SL(2,C_I) = Spin(1,3)$ spinor field, and an optical structure. We will also define a natural distribution on $\ML$ associated with this metric, and a natural lift of the Lorentzian structures and spinors to sub-pseudoriemannian structures on \ML. We also define natural holomorphic sections of $\ML$ on the base. 

In Section \ref{S:harm}, we will show that these sections are also harmonic maps, meaning that they are critical points of a variation problem and solve Euler-Lagrange equations. In the harmonic map case, these define the tension equation, involving the Weyl connection of the LCK or LCHK structures. We will show, by tensoring parallel spinor fields, that the tension equation is equivalent to a Dirac equation. Then, we will investigate the irreducible holonomy representations of the Weyl connection on \ML, in preparation for reducing the Dirac equations to a more familiar form.

In Section \ref{S:com} we will examine the induced equations on the holomorphic, anti-holomorphic, and hyperholomorphic tangent bundles. We will find that under reduction to natural distributions, these equations reduce to Dirac-Higgs equations for both the LCK and LCHK cases.

In Section \ref{S:tens}, we will take natural tensor products of these holomorphic and hyperholomorphic bundles, derive the Dirac-Higgs equations induced on them from the tension equation, and find that they are equivalent to the unbroken classical Dirac-Higgs equations of the Standard Model. We find an $SU(3)SU(2)U(1)$ interaction gauge group with:
\begin{itemize}
\item	the observed fermion pattern of $SU(3)$ color triplets and anti-triplets (quarks, anti-quarks), $SU(3)$ color singlets and anti-singlets (lepton, anti-leptons), $SU(2)$ weak doublets and singlets, linked correctly to chirality; 

\item	the observed pattern of $U(1)$ hypercharges for all fermions, both left and right, including fractional ones for quarks; 

\item	the observed three generations; and 

\item	an appropriate Higgs field coupling left and right spinor fields. 
\end{itemize}

In Section \ref{S:hyp} we will state our Òstrong hypothesis,Ó that physics is basically the geometry of the moduli space \ML. We propose that the differential geometry of \ML, defined by the structures and harmonic maps we investigated, is the classical theory of femionic and bosonic fields, i.e, the Standard Model unbroken classical field equations. 

In this section, we also investigate several related topics in the differential geometric picture. First, we show that the Yang-Mills equations, with fermionic currents, emerge naturally when the curvature of the Weyl connection is evaluated on the sub-pseudoriemannian distribution. (Unexpectedly, the Òcoupling constantsÓ of the Standard Model are replaced by an unbroken coupling field, analogous to the Higgs field, but derived from the Lorentzian metricÕs Levi-Civita connection form.)  Second, we examine the Einstein equations solved by our Lorentzian metric. Third, we suggest that the Standard Model Higgs coupling and generation Òmass matricesÓ are equivalent to our Higgs coupling structure. 

Finally, we propose that the algebraic geometry of $\ML$ is the quantum version of our classical field theory. It is defined by the sheaf equations that correspond to the Dirac-Higgs equations. The sheaf equations describe a class of parallel holomorphic maps that collectively define a holomorphic flow on $\ML$ and induce sheaf homomorphisms on the structure sheaf.  The new concept needed by QFT is a metric on sheaf sections, the Òvacuum expection valueÓ.

In Section \ref{S:fut}, we will outline a number of topics for future research and allow ourselves some speculation.

\section{Structures on \ML}\label{S:struct}

We will now derive \ML, the moduli space of stable $SL(2, C)$ holomorphic bundles over the hyperelliptic Hopf surface, \HL. Then we shall define two sets of structures on \ML: structures pulled back from \MLc, a compact moduli space that contains $\ML$ as an open set; and structures lifted from an open set of $\HP^1$, the quaternionic projective line, over which $\ML$ is naturally fibered.

\subsection{LCK and LCHK Structures on $\ML$}\label{S:pulled}

Let us define the Òhyper-ellipticÓ quaternionic Hopf surface, $\HL$, as
\[
\{ \Ho/\lambda^n ; \quad \lambda \colon q \to \lambda q ;  \quad  \lambda \in \R > 1, q \in \Hh \}
\]
with hyper-elliptic meaning only that it is obviously an elliptic surface in all complex structures.

There are two natural quaternionic structures induced on $X \in T_x \Ho \cong \Hh$ by left and right multiplication by $\{i, j, k\}$:
\begin{align*}
\I^- X &:= iX,     &     \J^- X &:= jX,           &   \K^- X &:= kX \\
\I^+ X &:= -Xi, & \J^+ X &:= -Xj, & \K^+ X &:= -Xk
\end{align*}
We will refer to these as the negative and positive quaternionic structures. Both are integrable for all $I = a\I + b\J +c\K; a^2+b^2+c^2 = 1$, thus \Ho has two hypercomplex structures. Under the $\lambda \colon q \to \lambda q$ homotheties,both hypercomplex structures are preserved, so they descend to \HL.  Thus \HL is a hypercomplex manifold with two hypercomplex structures.

\begin{notation}
Since the effect on real vectors $X$ of a change from negative to positive structures is simply to change $I \to -I, \forall I$, we shall refer to $\I^-$ as $\I$, and $\I^+$ as $-\I$, bearing in mind that for the positive operators the multiplication rule is $(-J)(-I) = (-K)$.
\end{notation}

Each complex structure $I = a\I + b\J +c\K$ on $T \Ho$ also reduces to $T\HL$ under the $\lambda$ homotheties, and each one induces an isomorphism between $\HL$ and a conventionally defined complex Hopf surface:
\[
\Ho/\lambda^n \cong \CIod/\lambda^n 
\]
Thus, for each $I$, there is a holomorphic elliptic fibration 
\[
\HL \to \CPI^1 
\]
with fiber isomorphic to the elliptic curve 
\[
\CIo/\lambda^n \cong \CI/\{2 \pi i \Z + \ln \lambda \Z \}
\]
Therefore, for each complex structure $I$, $\HL$ is elliptic. 

Braam and Hurtubise \cite{pb89}, followed by Lubke and Teleman \cite{ml95} constructed the moduli space of stable holomorphic $SL(2,C)$ bundles on an arbitrary elliptic Hopf surface as an open set of the moduli space, \MLc, of all (both stable and unstable) holomorphic $SL(2,C)$ bundles. In the case $c2 = 1$, the compact moduli space, $\MLc$, is especially straightforward: it is a manifold with a natural complex structure and locally conformal kaehler (LCK) metric. Thus, in our $c2 =1$ case this construction leads to an open embedding of $\ML$ into $\MLc$ and will let us pull back the LCK structure to the tangent bundle of \ML. Since our particular Hopf surface, \HL, is actually hypercomplex, we can also construct a hypercomplex structure on \MLc, and with it a naturally defined locally conformal hyperkaehler (LCHK) metric, and pull them also back to \ML.  Both metrics are homogeneous on \MLc, but under different groups, and both have associated standard connections. We shall keep track of both structures for the moment because they lead to different natural fibrations of \ML, one coming from the LCHK structure, and one coming from a choice of a particular complex structure and associated LCK metric.

\subsubsection{The Graph Technique for Constructing \ML}\label{S:graph}

Here is a brief sketch of the graph technique Braam and Hurtubise used to construct the moduli space of all stable holomorphic $SL(2,C)$ bundles on \HL. After specifying a complex structure, $I$, their technique starts with the elliptic (holomorphic) fibration of $\HL$ as 
\[
\TLI \to \HL \to \CPI^1 
\]
with fiber the elliptic curve $\TLI = \CI\backslash\{0\}/\lambda^n$. Then they reduce each bundle to a family of holomorphic bundles over the curve, and then employ the Atiyah classification \cite{ma57} of all holomorphic $SL(2,C)$ vector bundles on elliptic curves using $Pic(\TLI)$, the Picard group of the torus, which classifies all holomorphic line bundles on \TLI. The result is to canonically associate with each holomorphic $SL(2,C)$ bundle $E$ on $\HL$ a divisor (called a ÒgraphÓ) in the product space $P \times \CPI^1$, where $P = Pic^0(\TLI)/\langle i^0 \rangle$. ($Pic^0(\TLI)$ is the $c1 = 0$ component of $Pic(\TLI)$, and $\langle i^0 \rangle$ is the involution $L \to L^{-1}$ for all $L \in Pic^0(\TLI)$.) Since $P \cong \CPI^1$, the graph of $E$ is a divisor in $\CPI^1 \times \CPI^1$, and the set of all graphs is the linear system $|O(n,1)| \cong \CPI^{2n+1}$ where $n = c2(E)$. 

We are concerned with the $c2 = 1$ case, so the set of graphs for the bundles in our \MLc is $|O(1,1)| \cong \CP^3$. The complex structure on this $\CP^3$ comes from the original elliptic fibration of \HL, and thus from a particular complex structure $I$ associated with the isomorphism $H \cong \CI^2$. We will call this space of graphs $\CPI^3$. It is the projective space of $\CIof$. Each point in $\CPI^3$ is associated with a set of holomorphic bundles in \MLc and almost all of the points correspond to stable bundles. However, a real curve segment $\zeta \subset \CPI^3$ of points corresponds to unstable bundles.  The fibers of the graph map $Gr \colon \MLc \to \CPI^3$ are isomorphic to $\TLI$ in the case of both stable and unstable bundles \cite{ml95}. We can therefore define a complex structure on $\MLc$ by using the structures on $\CPI^3$ and $\TLI$, which both come from the original fibration and are the same as the complex structure on the $\CIof$ from which $\CPI^3$ is constructed by projectivization. (\MLc is topologically isomorphic to $S^1_\lambda \times S^7$.) $\ML$ is the open set of $\MLc$ defined by the stable fibers of the graph map.

The graph map is thus a holomorphic toric (principal) fibration over $\CPI^3$, depending on a particular complex structure of the original Hopf surface, with a natural action on each fiber by the torus group $\TLI$. This torus bundle is obviously covered by some holomorphic line bundle $O(n)$ on $\CPI^3$. (In fact, we shall learn below that the line bundle in question is $\mathcal{O}(-2)$.) The line bundle in turn is holomorphically covered by the tautological line bundle $O(-1)$, which is isomorphic (modulo the zero section) to the space $\CI^4\backslash\{0\}$. Thus $\MLc$ is covered by the kaehler manifold $(\CIof, e)$, where $e$ is the standard Euclidean metric.

\subsubsection{Locally Conformal Kaehler (LCK) Structure} \label{S:lck}

A Vaisman manifold, or generalized Hopf manifold, is basically a complex toric manifold covered by a kaehler manifold on which there is a discrete holomorphic homothetic action, i.e., a group of deck transformations. (See \cite{sd98}, ,\cite{lo09}, \cite{lo04}, \cite{mv03}, \cite{lo03}, and the notes therein, for more on Vaisman, LCK, and LCHK manifolds.)  The map $z \to \lambda z$ that defines the torus $\TLI$ is just such a holomorphic homothety, and thus \MLc is a compact Vaisman manifold.  These homothetic actions (dilations) are not isometries of the kaehler metric, so the Euclidean kaehler metric on $\CIof$ does not reduce to \MLc. But one may define another metric, $b$, on $\CIof$ that does reduce; it is called the Vaisman metric.  On open sets of \MLc, it is locally conformal to a kaehler metric.  The Vaisman metric has the properties 
\[
d\omega_b = \omega_b \wedge \theta, \quad d\theta = 0, \quad \lk{}{\theta} = 0 
\]
where $\omega$ is the kaehler form of $b$; the closed, parallel one-form $\theta$ is the Lee form; and $\lk{}{}$ is the Levi-Civita connection of $b$. Since $\MLc$ is not kaehler, $\lk{}{}$ does not preserve the complex structure, but a related connection, $\wk{}{}$, called the Weyl connection, does preserve it. The Weyl connection and its properties will be covered in much more detail in section \cite{S:tension}. We shall primarily follow Dragomir-Ornea, \cite{sd98}, in the discussion of the properties of LCK manifolds that follows. 

There is a natural association between a manifold with a toric fibration over $\CPI^3$, such as $\MLc$, and a 1-Sasakian structure $I \theta^\#$ on the  seven dimensional real sphere, $S^7$.  $\MLc$ fibers over $S^7$ with $S^1_\lambda$ fibers generated by the Lee field $\theta^\#$. $S^7$ then has its own Hopf fibration over $\CPI^3$, with $S^1$ fiber generated by the $I\theta^\#$ field \cite{cb08}.  In turn, a 1-Sasakian structure on $S^7$ implies a Vaisman structure on $\MLc$ itself.

The LCK metric has the general form
\[
b = \theta \otimes \theta + I \theta \otimes I \theta + k
\]
where $k$ is kaehler on the $dim_\C = 3$ distribution 
\[
T := ker (\theta + I\theta)
\] 
The Lee field $\theta^\#$ and the complex structure $I$ define another distribution that is vertical to the toric fibers of the graph map and also $b$-orthogonal to $T$:
\[
V := \langle \theta^\#, I\theta^\# \rangle
\]
Thus the graph map induces a natural mapping of $T$ onto $T\CPI^3$, and $k$ is the lift to $T$ of the Fubini-Study metric on $T\CPI^3$. In addition, $dI\theta$ is the lift of the kaehler form of the Fubini-Study metric, which is the curvature the induced connection on the canonical complex line bundle $K = O(-4)$. Thus the LCK metric $b$ on \MLc and the principal toric fibration $\MLc \to \CPI^3$ induces a a kaehler metric $k$ on $T\CPI^3$ and an associated $U(1)$-connection, $iI\theta$, on the line bundles $\mathcal{O}(n)$ over $\CPI^3$.  The form $\theta$ itself is a flat connection on the trivial ``weight'' bundle on $\CPI^3$, since the curvature of $\theta$ is $d\theta = 0$. Both the complex structure, $I$, and the LCK metric, $b$, can be pulled back to $\ML$ from $\MLc$.

\subsubsection{Locally Conformal Hyperkaehler (LCHK) Structure}\label{S:lchk}

If the graph map of $\ML$ (and \MLc) is restricted to an individual projective line $\CPI^1$ in $\CPI^3$, the result is isomorphic to the original Hopf surface \cite{ml95}, that is, to $\Ho/\lambda^n$, which has a natural hypercomplex structure $(\I, \J, \K)$ defined by the left multiplication by the quaternions $(i, j, k)$.  Thus there is a natural principal fibration of $\MLc$ over $\HP^1$, the quaternionic projective line defined by left quaternionic multiplication, with fiber isomorphic to the original Hopf-surface group manifold and a natural vertical distribution on $\MLc$ defined by
\[
S := \langle \theta^\#, \I\theta^\#, \J\theta^\#, \K\theta^\# \rangle 
\]

This principal fibration allows us to extend the 1-Sasakian structure on $S^7$ to a 3-Sasakian structure. The 3-Sasakian structure is then is canonically associated with a locally conformal hyperkaeher (LCHK) structure on $\MLc$, which we will call $(\MLc, \I,\J,\K, h)$, where $h$ is the LCHK equivalent of $b$. Like $\HL$, this manifold is hypercomplex, and it is covered by the hyperkaehler manifold $(\Hod, \I,\J,\K, e)$, where again $e$ is the Euclidean metric. (Clearly homothetic action by real $\lambda$ preserves this hypercomplex structure and the LCHK metric $h$, but not the Euclidean metric $e$.) 

We shall call the group of this principal bundle $S^1_\lambda Sp(1)^-$. This fibration of \MLc is the so-called the (negative) ÒSwann fibrationÓ of an LCHK manifold over the quaternionic manifold $\HP^1$.) It is isomorphic to a compactification of the negative spinor bundle $\$^-$ for the induced riemannian structure on $\HP^1$ and is covered by the tautological \Hh-line bundle on $\HP^1$ \cite{as91}. (The tautological \Hh-line bundle itself is sometimes called the Òassociated bundle to a quaternionic kaehler manifold,Ó as well as the negative spinor bundle.) The compatification is induced by the $\lambda^n$ homotheties, and the covering map is holomorphic for all $I$. After a choice of $I$, this bundle fibers over $\CPI^3$ and that fibration is isomorphic to $\mathcal{O}(-2)$. Thus \MLc is a $\lambda^n$-compactified $\mathcal{O}(-2)$.

The LCHK metric $h$ on $\MLc$ has the following form:
\[
h = \theta \otimes \theta +\I\theta \otimes \I\theta +  \J\theta \otimes \J\theta + \K\theta \otimes \K\theta + q
\]
where $q$ is quaternionic kaehler on the $dim_\Hh = 1$ distribution 
\[
D = ker (\theta +\I\theta +\J\theta +\K\theta) 
\]
which under the Swann fibration is mapped onto $T\HP^1$. Thus the induced riemannian metric on $\HP^1$ is $q$.

Associated with the LCHK metric on $\MLc$ is a Levi-Civita connection $\lhk{}{}$ and a Weyl connection $\whk{}{}$. The Weyl connection preserves all the complex structures in the hypercomplex structure. It will also be discussed in section 3.

So we are left with two sets of structures that pull back to $\ML$ from $\MLc$, an LCHK structure and a LCK structure associated with a particular complex structure $I$. We also have associated with these structures Levi-Civita connections and Weyl connections on the tangent bundle. The Levi-Civita connections preserve the metrics, while the Weyl connections preserve complex or hypercomplex structures. With these pulled-back structures, $\ML$ becomes a (non-compact) LCK and LCHK manifold. It is clearly covered by a complex line bundle on $\CP^3_{Io} := \CPI^3\backslash\{\zeta\}$, the open set of stable graphs in $\CPI^3$. Since $\ML$ is Vaisman, the total space of that line bundle is a kaehler manifold.

\subsubsection{Quaternionic Twistor Fibration}\label{S:twist}

We have defined two fibrations of \MLc: a toric fibration over $\CPI^3$, isomorphic to a compactified $\mathcal{O}(-2)$, and a (negative) Swann fibration over $\HP^1$, isomorphic to a compactified tautological bundle. Similarly, there are two fibrations of $\ML$ over open subsets of $\CPI^3$ and $HP^1$: over $\CP^3_{Io}$ and over $\HP^1_o$.

The induced fibrations 
\[
\CPI^3 \to \HP^1 \quad  \CP^3_{Io} \to \HP^1_o 
\]
of the complex contact manifold $CPI^3$ over the quaternionic manifold $\HP^1$ is the "quaternionic twistor fibration" associated to the Swann negative fibration by a choice of complex structure $I$. (See Salamon \cite{ss82} for a clear exposition of the quaternionic twistor bundle.)  The twistor fibration will play a key role in the next section.

\subsection{Sub-Pseudoriemannian and Optical Structures, Holomorphic Sections}

Let us now look at more closely at the effect of a few group actions on $\ML$ induced by corresponding natural group actions on $\HL$.  First, as we said earlier, $\HL$ is a group manifold, and there is a natural action of $\HL$ on itself defined by the action of $\Ho = exp(\Hh)$ on itself by left quaternionic multiplication $q_1 q$. This action on $\HL$ induces an action on $\ML$ that is vertical on the fibers of the negative Swann bundle and generated by the vector fields $\langle\theta^\#, \I\theta^\#, \J\theta^\#, \K\theta^\#\rangle$. This action permutes the complex structures $\I$, $\J$, and $\K$ \cite{dj92} and induces a natural principal $S^1_\lambda Sp(1)^-$ bundle structure for the Swann bundle, $\ML$, fibered over $HP^1_o$. Similarly, right multiplication on $\HL$ induces an action on the fibers of the positive Swann bundle (itself derived from the compactification of the right spinor bundle). 

Next, let us consider $(\HL, I)$ as a complex manifold, where $I$ is the complex structure on $T \HL$ induced by (left multiplication by) an imaginary unit quaternion of \Hh. The automorphism group of $(\HL, I)$ (that is, the group that preserves $I$) is equal to $GL(2,C_I)/\lambda^n$ \cite{pb89}. It induces a similar $GL(2,C_I)/\lambda^n$ action on $\ML$ that preserves the $\HL$ fibers of the Swann bundle and one of their complex structures, namely $I$. This group acts in a familiar way on each fiber, as $\TLI SL(2,C_I)$, where $\TLI$ is just the torus group in the toric fibration over $\CP^3_{Io}$.  

Thus the holomorphic automorphism group of $\CP^3_{Io}$ is $SL(2,C_I)$, which acts on the quaternionic twistor fibration $\CP^3_{Io} \to \HP^1_o$ via Moebius transformations of the $\CP^1$ fibers. There is also a conjugate action of $SL(2, C_I)$ on $\CPI^1$ fibers via the conjugate Moebius action. This is equivalent to the standard Moebius action of $SL(2, C_{-I})$, that is to say of a reversed complex structure. These two Moebius actions on $\CPI^1$ fibers are projectivizations of the left and right spinor representations of $SL(2,C_I)$. We shall see that they also correspond to the negative and positive Swann fibrations, and thus to left and right quaternionic multiplication. (The isomorphism $Sp(1)^- \cong SU(2)_I \hookrightarrow SL(2, C_I)$, induced by a selection of the complex structure $I$, leads to the identification of the negative quaternionic twistor bundle, with the projectivization of the left $SL(2, C_I)$ spinor bundle. A similar isomorphism,  $Sp(1)^+ \cong SU(2)_{-I} \hookrightarrow SL(2, C_{-I})$ identifies the positive twistor and projectivized right spinor fibrations.

\subsubsection{Lorentzian Metric on ${\HP^1_o}$}\label{S:metric}

Let us focus on this $SL(2, C_I)$ action on the negative quaternionic twistor bundle, 
the projectivization of the $Sp(1)^-$-spinor bundle,  which Swann identifies with the tautological $\Hh$-line bundle on $\HP^1$. The $\C_I^2$ fiber of this spinor bundle holds the spinor representation of the $SU(2)_I \cong Sp(1)^-$ principal bundle over $\HP^1_o$ that we introduced in \ref{S:lchk}. 

Normally, e.g. by Salamon, \cite{ss82}, this spinor bundle, $\$^-$, is a chosen as the negative half-spinor representation of the Clifford bundle $Cl(\HP^1_o, q)$, defined by a riemannian metric $q$ induced on $\HP^1_o$ by the LCHK metric on $\Hod$ and an $Sp(1)^- Sp(1)^+ \cong Spin(4)$ spin frame bundle. (Actually, Salamon uses the positive spinor bundle) In this section, however, we will use the Moebius $SL(2,C_I)$ action on the twistor fibers to interpret that same bundle as a left or right half-spinor representation, $\$^L$ or $\$^R$, of the Clifford algebra $Cl(HP^1_o, g)$, defined by a Lorentzian metric $g$ and the $SL(2, C_I) \cong Spin(3,1)$ spin frame bundle.  In this set up, we have two possible $Spin(3,1)$ spinor bundles: $\$^L$ and $\$^R$.

A Lorentzian metric $g$ on $\HP^1_o$ defines a Levi-Civita connection, $\lcl{}{}$, whose associated spin connection, $\spl{}{}$, acts on sections of the spinor bundle, $\$$. It also acts on chiral (Òhalf-spinorÓ) sections of the spinor bundle, $\$ = \$^L + \$^R$, that is, those that fall entirely within $\$^L$ or $\$^R$. We have chosen $\$^L$ to be the (non-compact) negative spinor bundle, $\$^-$, i.e., the tautological bundle of $\HP$ determined by the projective fibration $\Hod \to \HP^1$, which means it covers \ML. 

So far, so good; but can we construct $SL(2,C_I)$ metrics $g$ canonically? In fact, from the geometry of Lorentzian manifolds we know that we can define a canonical class of Lorentzian metrics on $\HP^1_o$ that uniquely determine an associated spinor field and vector field. Locally, as we shall see in the next subsection, these metrics also determine an optical structure on $\HP^1_o$ and a class of local holomorphic maps to \ML. Using these structures and maps, we can link the $g$ metrics to $(\ML, b)$ in a very different way than is provided by the usual construction of $q$.

Let us require that the metric $g$ support a non-zero, parallel parallel Dirac spinor, $\Ps \in \$\backslash\{0\}$, of the Clifford algebra of the metric. These metrics have been studied extensively, and are known as ÒBrinkmann metricsÓ or Òpp-wave metrics,Ó see \cite{rb91}. Canonically associated to the parallel spinor is a parallel non-zero null vector field called the ÒDirac currentÓ: 
\[
N_g : = \langle \Ps, e^{\mu} \cdot \Ps \rangle e^{\mu}
\]  
where $\langle, \rangle$ is an induced scalar product on \$, $\cdot$ is the Clifford multiplication, and $e^{\mu}$ are the frame vectors. The Dirac current field is Killing and integrates to a null directed flow. 

The parallel spinor fields are chiral, that is, sections of either $\$^L$ or $\$^R$. Sections of both chiralities exist for any such metric, each represented by a half-spinor field; and, most importantly, the spinor field is unique up to a constant. Let us call the unique parallel non-zero left spinor field $\PsL$ and the unique parallel non-zero right spinor field $\PsR$. (In the $(3,1)$ signature, these fields can be represented by identical $CI^2$ sections, that is, pure $SU(2)$ spinor fields, \cite{rm02}.) 
 
The spinor fields and the Dirac current are canonically associated to the metric by the parallelism condition: 
\begin{align*}
 \spl{}{\Ps} &= 0,   &  \spl{}{\PsL} &= 0, & \spl{}{\PsR} &= 0 \\
\lcl{}{N_g} &= 0   &   ||N_g||_g &= 0            
\end{align*}

\subsubsection{Optical Structure}\label{S:optic}

Clearly a non-zero parallel left spinor field, $\PsL$, determines a global section of the projective bundle $\mathbb{P}(\$^L)$. We will call these sections of the projective bundle, $\PsTL$. As we have defined it, $\mathbb{P}(\$^L)$ is identical to the quaternionic twistor bundle $\mathbb{P}(\$^-) =: Z(\HP^1_o, g)$, so $\PsTL$ is a section of the left Lorentzian twistor bundle. (Consult Leitner \cite{fl98}, for a discussion of Lorentzian twistor space realized as projective Lorentz left and right spinors.)

$Z(\HP^1_o, q)$ is the set of almost complex structures on the \emph{riemannian} manifold $(\HP^1_o, q)$, and thus a section $\phi$ of \emph{this} twistor bundle would define an $q$-compatible almost complex structure $I_\phi$ on $(\HP^1_o, q)$ and thus an almost hermitian structure. ($I_\phi$ is defined at each point $\phi(x) \in Z(\HP^1_o, q)$ by the action of the almost complex structure $I$ on the $\nabla_q^{LC}$-horizontal subspace of the tangent space of the twistor bundle $Z(\HP^1_o, q)$. Since $I$ is actually integrable on $Z(\HP^1_o, q)$, $q$ must be anti-self-dual \cite{ss82}. If $I_\phi$ is parallel with respect to $\nabla_q^{LC}$, then $(\HP^1_o, q)$ is kaehler.

But since $g$ is Lorentzian instead of riemannian, a section $\PsTL$ of the $Spin(3,1)$ twistor space $\mathbb{P}(\$^L) =: Z(\HP^1_o, g)$ instead determines a \emph{optical} structure $O_\psi$, with the Lorenzian metric. (Optical structures on Lorentzian manifolds are very similar to hermitian structures on riemannian manifolds. See \cite{fl98}, \cite{fl01}, and \cite{pn96}, for more discussion of optical structures associated with Lorentzian manifolds via the Lorentzian twistor fibration.) Since $\PsTL$ is parallel with respect to $\nabla_g^{LC}$, so is $O_\psi$. (This is equivalent to the statement that $(r, I_\phi)$ is kaehler in the hermitian case.) The riemannian requirement that $q$ be anti-self-dual in the integrable case is replaced by a requirement that $g$ be conformally flat \cite{pn96}. So we shall require that our metric support non-zero parallel spinors and be conformally flat. (Pp-wave metrics meet that requirement.)

\subsubsection{Local Holomorphic Sections}\label{S:sect}

Since $\PsL$ is non-zero, it also defines a section $\PL$ of $\ML$ itself, considered as the negative Swann  bundle, covered by $\$^- = \$^L$. The differential $d\PL$ sends $T \HP^1_o$ into $T\ML$.

Our $\PL$ defines an entire equivalence class, $| \PL |$, of holomorphic sections of \ML, with their 
differentials sending $T \HP^1_o$ into $T\ML$. The elements of the class are $\gamma \circ \PL
$ where $\gamma$ is an element of the group $S^1_\lambda SU(2)_I$.  ($SU(2)_I \hookrightarrow SL(2,C_I))$. $SU(2)_I$ acts on the fibers of the Lorentz twistor bundle with isotropy group $U(1)_I$. The isotropy action and the $S^1_\lambda$ action combine to generate the toric action $\TLI$.) Each element of the class $| \PL |$ is essentially the same as the original, because each is related to $\PsL$ by either scalar multiplication or a rotation of the basis of $\$^L$ which preserves the spinor inner product. The distribution of $\ML$ created by $| \PL |$, we will call $E$. 

Lift the pseudoriemannian structure $g$ to the distribution $E$. (By definition, $E_z$ at any point $z \in \ML$ is isomorphic to the tangent space at $\pi^- (z)$ of $\HP^1_o$ under the differential of the negative Swann fibration $d\pi^-$.) If $g$ were positive definite, this structure on $\ML$ would be called Òsub-riemannian,Ó but in our case it is Òsub-pseudoriemannian.Ó With this structure we can construct a lifted Levi-Civita (sub)-connection, which we will also call $\lcl{}{}$ and a lifted spin (sub)-connection $\spl{}{}$, which will operate on lifted (sub)-spinors. See \cite{rm02} for a description of sub-riemannian geometry (also called Carnot-Catheodory geometry).

By the definition of the twistor construction of the optical structure on $HP^1_o$, and since $g$ is conformally flat, the differential, $d\PL$, of each $\PL$ in the class, also sends $O_\psi$ on $HP^1_o$ into a lifted optical structure of the image leaf of $E$, which is just the complex structure $I$ on \ML. Thus, $I \circ d\PL = dPL \circ O_\psi$, which is the definition of a holomorphic map from an optical manifold to a complex manifold \cite{fl01}. Thus, $\PL$ is a holomorphic map $\HP^1_o \to \ML$.

\section{Harmonic Maps and Dirac Equations}\label{S:harm}

Since $\PL$ is holomorphic and, since $(\HP^1_o, g, O_\psi)$ is the Lorentzian equivalent of 
kaehler and $(\ML, b, I)$ is LCK, $\PL$ is also harmonic \cite{sd98}. Harmonic maps are 
often called ``non-linear sigma models'' in the physics literature. Needless to say, both the 
mathematical literature on harmonic maps, initially notably by Eells and his collaborators 
(e.g., \cite{je83}), and the physics literature on non-linear sigma models, e.g, \cite{sk00}, is vast. Our
 harmonic maps $\PL$ are non-linear sigma models from a 4-dimensional Lorentzian 
manifold to an 8-dimensional non-compact moduli space with both LCK and LCHK structures.

We have defined two manifolds, $(\HP^1_o, g, O_\psi)$ and $(\ML; b, I; h, \I, \J, \K)$ and harmonic 
maps between them. We will now use this setup to derive the first of three important sets of 
equations of the classical (unbroken) Standard Model of physics: the Dirac-Higgs equations 
for fermions.  The Yang-Mills equations, with currents along with the equations for the Higgs 
field, and the Einstein equations will be deferred until later sections. 
      
The whole process begins with the definition of a harmonic map by the Euler-Lagrange 
equations associated with the critical points the integral of the map density, in the same way 
that the classical Standard Model equations are the Euler-Lagrange equations for the critical 
points of the integral of the Lagrangian density. Unlike the conventional formulation of the 
Lagrangian density, however, the usual expression of the energy density of the map is quite 
terse, with most of the information in the structures on the manifolds, such as the metrics. So 
while in physics it is usually sufficient to describe the Lagrangian, which can be incredibly 
elaborate (see \cite{js12} for the current best description of the Standard Model Lagrangian), in 
our theory it will be necessary to work out the equations in detail to show their equivalence to 
the physical ones.
    
\subsection{Tension Equation of Harmonic Maps} \label{S:tension}

Harmonic maps $f: (M, g) \to (N, h)$ are smooth maps at the critical points in the map's ``energy'' 
functional: 
 \[
 E(f) = \int_M e(f) dM  \quad           e(f) := ||df||^2_{gh} 
 \]
where the norm involves the both metrics $g$ and $h$, and $e(f)$ is usually called the ``energy 
density'' in the mathematics literature. (The term ``energy'' annoys physicists, since it means 
something a little different in physics. Their preferred terms are ``action'' for the integral, and 
``Lagrangian density'' for $e(f)$.) The Euler-Lagrange equations of this variation problem define the 
harmonic maps as those with vanishing tension field: 
 \[     
\tau(f) := tr_g(\nabla_h^{LC} df) = 0
 \]    
where the Levi-Civita connection of the target manifold is used, and the trace is taken using the metric 
of the source manifold \cite{je83}. We shall define the trace shortly.

Harmonic maps that arise from holomorphic maps into kaehler manifold targets (such as is the case for 
our spinor section $\PsL$, which map to the kaehler cover of $\ML$), use the Levi-Civita connection of 
the kaehler metric $e$:
\[      
	\tau(\PsL) := tr_g(\nabla_e^{LC} d\PsL) = 0
\]
In the case of non-kaehler, but LCK, metrics, such as $(\ML, b, I)$, the Levi-Civita connection is simply 
replaced in the tension equation by $\nabla_b^w$, the Weyl connection of the LCK metric, where
\[
	\nabla_b^w = \nabla_b^{LC} + (1/2)(b \otimes \theta^\# - \theta \otimes Id - Id \otimes \theta)
\]
and $\nabla_b^{LC}$ is the ordinary Levi-Civita connection of the metric $b$, which does not preserve 
the complex structure $I$ although it does of course preserve the metric $b$ \cite{sd98}. ($Id$ is the identity operator on $T \ML$.) Thus, on vector fields $X$ and $Y$ over $\ML$, $\nabla_b^w$ acts as follows:
\[
	\wk{X}{Y}  = \nabla_{b X}^{LC} Y + (1/2)(b(X,Y)\theta^\# - \theta(X)Y - \theta(Y)X)
\]
So the tension equation for our harmonic maps into $(ML, b, I)$ is
\[
      \tau(\PL) := tr_g(\nabla_b^w d\PL) = 0
\]      

This trace equation is written as the trace (over the source metric) of a (vector-valued) two-form. But 
$d\PL$ is a one-form on the source, while $\wk{}{}$ is a one form on the target. So we need either to 
pull back $\wk{}{}$ and the target tangent bundle to $\HP^1_o$ or lift $d\PL$ and $g$ to the target. 
Normally, the approach for arbitrary harmonic maps between manifolds is to use pullbacks. Since our 
maps are sections of bundles with distributions $E$, we are able to pursue 
lifts. Lifting the form is fairly straightforward. On any $X = d\PL(e)$ in a leaf of $E$ (for 
some $e \in \Gamma(T \HP^1)$), the lift, $h^*d\PL$, acts simply as the identity: $(h^*d\PL)
(X) = Id(X) = X = d\PL(e)$. But for this to work we need to have an image of $e$ under $d\PL$ in each 
leaf of the distribution.
  
Our equivalence class, $| \PL |$, of holomorphic sections, with their differentials targeting each leaf of the distribution $E$, \ref{S:metric}, provides that image. We will then read 
the tension equation as taking the trace of $\wk{}{d\PL}$ over all elements of the equivalence class $| d
\PL |$, with the trace taken using the lifted metric on each leaf of $E$. We will therefore read that 
equation as the evaluation of an $\ML$ two-form on the symmetric two-vector $g^*$ on $\ML$, which is 
defined as the ÒcometricÓ in the sub-pseudoriemannian structure \cite{rm02}. Thus, our tension equation 
links the LCK structure on $\ML$ with the sub-pseudoriemannian structure. On $\ML$ it has the form
 \[     
      \wk{}{(g^*)}  -  \lcl{}{(g^*)} = 0
 \]     
where $\lcl{}{}$ is now the lifted Levi-Civita connection of $g$.  Essentially the tension 
equation says that the evaluation of the Weyl connection of $b$ on the lifted two-vector $g^*$ is 
equal to that same evaluation of $g^*$ by the lifted Levi-Civita connection of $g$.  If the target were 
kaehler, it would be an equation connecting two Levi-Civita connections.

      By the definition of the trace, this equation is
\[      
	\sum_{\mu\nu} g_{\mu\nu} [\wk{}{(X^\mu , X^\nu)}  - \lcl{}{(X^\mu, X^\nu)}]  = 0
\]
where $X^\mu$ is the image under the $d\PL$ mapping of a frame field $e^\mu$ on $\HP^1_o$. We 
will normally use the Einstein summation convention and write this as
\[
	\wk{X^\mu}{(X^\mu)}  -  \lcl{X^\mu}{(X^\mu)} = 0 	
\]
  
\subsection{Dirac Equation Equivalent} \label{S:dirac}

We will now take a step that will be a recurring feature of our process throughout the rest of this paper, 
namely, to take a version of the tension equation an rewrite it as a Dirac equation by tensoring $X^\mu
$, or its equivalents, with a lifted parallel spinor field (either $\Ps$, $\PsL$ or $\PsL$).

\subsubsection{Jordan Multiplication and the Tensor Product}\label{S:jordan}

We can put the tension equation into the form of a Dirac equation by examining the structures we have 
defined on $\HP^1_o$ in the previous section. Along with the metric $g$, we defined its Clifford 
algebra $Cl(\HP^1_o, g)$ Also associated with the Lorentzian metric is  a ``Jordan algebra of the Clifford type,'' 
$J(\HP^1_o, g)$, (also called a ``spin factor'') \cite{hu87}. At each point of $\HP^1_o$ this algebra is generated by a 
timelike basis vector $s^0$ and three orthonormal space-like basis vectors $\{s^j, j = 1, 2, 3\}$. This algebra is isomorphic to the matrix algebra of $2 \times 2$  Hermitian matrices, with the (symmetric) algebra product defined as 
\[
s^\mu \cdot s^\nu = \frac{1}{2}(s^\mu s^\nu + s^\nu s^\mu)
\]
One convenient representation of the basis is 
\[
s^0  := \sigma^0 = Id , \  s^j := \sigma^j,
\]
the identity matrix and  the Pauli matrices. Clearly $s^\mu = s^0 \cdot s^\mu$, since $s^0 = Id$.

As a Clifford-compatible Jordan algebra, it also has a Jordan multiplication of  Weyl spinors in $\$^L$ or $\$^R$ that is the same as the Clifford multiplication of Weyl spinors by $Cl(s^j)$. This Clifford algebra is the even sub-algebra $Cl^0(\HP^1_o, g)$, of  $Cl(\HP^1_o, g)$, and is generated by the space-like basis vectors, $\{s^j\}$. This mapping allows us to define a representation of the Jordan multiplication of Dirac spinors in $\$$. The basis vectors of $J(\HP^1_o,g)$ map into elements of $Cl^0(\HP^1_o, g)$, and hence of  $Cl(\HP^1_o, g)$. One straightforward representation of this mapping, using the Pauli-matrix, $\sig$ form of the $J(M,g)$ basis, and the $\gam$ representation of the $Cl(\HP^1_o, g)$ generators, is
\[
\sig \to \gamma^{\mu 0} := \gam \cdot \gamma^0 = (Id, \gamma^j \cdot \gamma^0)
\]
(Notice that the image of $\sig$ in $Cl(\HP^1_o, g)$, consists of the \emph{bivectors} $\gamma^{j 0} := \gamma^j \cdot \gamma^0$, plus the \emph{scalar} $\gamma^{0 0} := \gamma^0 \cdot \gamma^0$.) Jordan multiplication of spinors by vectors maps into Clifford multiplication of spinors by the bivectors and scalars. All these Jordan and Clifford structures lift to the distribution $E$.

Moving from the basis vectors of the tangent space at a single point to the frame fields $e^\mu$ of $T\HP^1_o$, using Jordan multiplication of vectors by vectors we can write the frame fields as  
\begin{equation}\label{E:sigmult}     
      e^\mu = e^0 \cdot e^\mu
\end{equation}     
where Ò$\cdot$Ó represents Jordan multiplication, and $e^0$ is the unit timelike vector field. The lifted versions of these frame fields, $X^\mu := d\PL(e^\mu)$, can also be Jordan multiplied by lifted vector fields.

\begin{notation}
Beginning at this point, we will use $\sig$ to represent lifted frame fields of the lifted Jordan bundle $J(M,g)$, when they act on vector fields by Jordan multiplication or on Weyl spinors by Jordan multiplication. Let us write $X^\mu = X^0 \cdot \sig$ for the lifted version of equation \ref{E:sigmult}.

Similarly, we will use $\gam$ to represent the lifted generator fields of the lifted Clifford bundle $Cl(\HP^1_o, g)$, when they act on Dirac spinor fields by Jordan/Clifford multiplication. Thus we will write the lifted Jordan multiplication of Dirac spinor fields by these lifted generator fields as $\gam \cdot \Ps$. \end{notation}

Now let us construct the lifted tensor product bundle $d\PL(T\HP^1_o) \xj \$$, where $\$$ also 
represents the lifted version of $\$$ and $\xj$ means the tensor product over the Jordan algebra. 
Sections of this bundle are vector-valued $\$$-spinor fields. We can also construct a tensor product 
connection on the lifted product bundle as
\[      
      	\wksl{}{} := \wk{}{} \xj \spl{}{}
\]      
and a section that is the tensor product of the lifted frame fields with the lifted parallel spinor field 
\[
      X^\mu \xj \Ps 
\]
We can also apply Jordan multiplication within the tensor product, since it it is over the Jordan algebra:
\[
X^\mu \xj \Ps = (X^0 \cdot \sig) \xj \Ps = X^0 \xj (\gam \cdot \gamma^0 \cdot \Ps) = X^0 \xj (\gam \cdot \Pst)
\]
where $\Pst = \gamma^0 \cdot \Ps$ is the ``inverted'' $\Ps$ field, that is, the $\PsL$ and $\PsR$ components are exchanged. Notice that Jordan multiplication of spinors sends $\$^L$ into $\$^R$ and vice versa (thus, $\gamma^0 \colon \PsL \to \PsR$).    
In the chiral representation
\[
\Ps = \begin{pmatrix} \PsR \\ \PsL \end{pmatrix} \quad \Pst = \begin{pmatrix}  \PsL \\ \PsR \end{pmatrix}
\]

\subsubsection{The Tension Equation as Dirac Equation}\label{S:tensdir}

Since the spinor field $\Ps$ is $\lcl{}{}$-parallel, we can write our tension equation as
\[      
\wksl{X^\mu}{(X^0  \xj \gam \cdot  \Pst)} - (\lcl{X^\mu}{X^\mu}) \xj \Ps = 0
\]
Clearly, 
\begin{align*}
\wksl{X^\mu}{(X^0 \xj \gam \cdot \Pst)} &= \gam \cdot (\wksl{X^\mu}{(X^0 \xj \Pst)}) + X^0 \xj (\lcl{X^\mu}{X^\mu \cdot  X^0 \cdot \Ps)}\\
X^0 \xj (\lcl{X^\mu}{X^\mu} \cdot X^0 \cdot \Ps) &= X^0 \cdot (\lcl{X^\mu}{X^\mu}) \xj \Ps = \lcl{X^\mu}{X^\mu} \xj \Ps
\end{align*}
So, after expanding, the two terms in the tension equation involving $\lcl{}{}$ cancel, leaving
\[
	\gam \cdot \wksl{X^\mu}{(X^0 \xj \Pst)} = 0
\]
which is just the Dirac equation for the connection $\wksl{}{}$ and the vector-valued spinor field $X^0 \xj \Pst$. 

Since the ordering of $\PsL$ and $\PsR$ was arbitrary, and the parallel spinors are chiral in any case, we could have started with our tensor product defined as $X^\mu \xj \Pst$, and arrived at the Dirac equation:
\[
	\gam \cdot \wksl{X^\mu}{(X^0 \xj \Ps)} = 0
\]

\begin{notation}
To avoid carrying the tilde marker so much, we will use the Dirac equation form with \Ps instead of \Pst.  Also, since \Ps (and \Pst) are \spl{}{} parallel, we will drop the $g$ subscript and the $s$ superscript from the connections \wksl{}{}, \wkslc{}{}, and \whksl{}{}. (I realize that the passion for avoiding extra indices is seizing me rather late in the day.)
\end{notation}

Our Dirac operator in this equation is 
\[
\gam \cdot \wk{X^\mu}{}, 
\]
where the Jordan/Clifford multiplication acts on the spinor components, while the Weyl connection acts on the 
vector components. 

If we had started with the LCHK manifold $(\ML, h, \I,\J,\K)$ as the target, but selected the same 
complex structure $I$ in the two-sphere of complex structures 
\[      
S^2 := a\I + b\J +c\K \quad a^2+b^2+c^2 = 1
\]
on $\ML$ in the definition of our holomorphic maps, we would arrive at a similar equation, but 
using the Weyl connection for $h$:
\[    
 	\gam \cdot \whk{X^\mu}{(X^0 \xj \Ps)} = 0
\]

To summarize: for harmonic maps
\[      
  \PL \colon (\HP^1_o, g) \to (\ML, b) \colon (\HP^1_o, g) \to (\ML, h)
\]      
the lifted tension equations for the LCK and LCHK cases are:
\begin{align*}
      \wk{X^\mu}{X^\mu} - \lcl{X^\mu}{X^\mu} &= 0 \\     
      \whk{X^\mu}{X^\mu} - \lcl{X^\mu}{X^\mu} &= 0
\end{align*}      
where $X^\mu = d\PL(e^u)$.  The equivalent Dirac equations for LCK and LCHK vector-valued 
spinor fields are
\begin{align*}
	\gam \cdot \wk{X^\mu}{(X^0 \xj \Ps)} &= 0\\
	\gam \cdot \whk{X^\mu}{(X^0 \xj \Ps)} &= 0
\end{align*}

Finally, since our parallel spinor are \emph{chiral}, that is, entirely in $\$^L$, or $\$^R$, it is sometimes convenient to write pairs of equations left or right parallel spinor fields, $\PsL$ and $\PsR$. In the chiral representation of the $\gamma^\mu$ frame fields, 
\begin{equation}\label{E:gamma}
\gamma^0 = \begin{pmatrix} 0  \quad \sigma^0 \\ \sigma^0  \quad \  0\end{pmatrix}, \quad
\gamma^j = \begin{pmatrix} 0 \quad \sigma^j \\ -\sigma^j \quad 0\end{pmatrix}
\end{equation}
at a point $x$, the Jordan multiplication by frame fields are different on the left and right Weyl spinors:
\begin{align*}
\sig \cdot \PsL &:= (\sigma^0, \sigma^j) \cdot \PsL \\
\sigr \cdot \PsR &:= (\sigma^0, -\sigma^j) \cdot \PsR
\end{align*}
So the independent left and right spinor field ``Dirac-Weyl'' equations are:
\begin{align*}
	\sig \cdot \wk{X^\mu}{(X^0 \xj \PsL)} &= 0\\
	\sigr \cdot \wk{X^\mu}{(X^0 \xj \PsR)} &= 0\\
       \sig \cdot \whk{X^\mu}{(X^0 \xj \PsL)} &= 0\\
	\sigr \cdot \whk{X^\mu}{(X^0 \xj \PsR)} &= 0
\end{align*}

From time to time in the rest of the paper we will perform this process of taking a form of the tension equation 
and writing it as an equivalent Dirac equation, after tensoring a vector field with a lifted parallel spinor field.  
      
The next few sections will deal with reducing these equations to a more transparent form on various 
distributions tied to the holonomy reduction of the Weyl connection component of our Dirac equations 
(the component acting on the vector field).

\subsection{Holonomy Representations of the Weyl Connection}\label{S:holon}

In this section we will show that the Weyl connection holonomy representation reduces on $T\ML$, both in the LCK case and in the LCHK case. In the LCK case we have an $SU(3)$ complex triplet 
and an $SU(3)$ trivial complex singlet, while in the LCHK case we have an $Sp(1)$ 
non-trivial  quaternionic singlet and an $Sp(1)$ trivial  quaternionic singlet. 
      
We know that the $(b, I)$ and $(h, \I,\J,\K,)$ structures on $T\ML$ are LCK and LCHK, respectively, and 
have Weyl connections with holonomy equal to the holonomy of the Levi-Civita connections of the kaehler 
and hyperkaehler metrics of covering spaces \cite{lo09}. Since the holonomy group of the Levi-Civita 
connection of a symmetric space is just the isotropy group, to find holonomy we will first look at the 
homogeneity properties of the complex and hypercomplex structures and the metrics on the kaehler covering 
spaces $\CIof$ and $\Hod$. The homogeneity properties of  $\CIof$ are well known; we will 
follow Joyce's discussion \cite{dj92} of homogeneous hypercomplex structures. 
      
\subsubsection{LCK Weyl Holonomy Reduction}\label{S:lckholon}

$(\CIof, e, I)$ is clearly orientable, and, after selecting an orientation, the isometry group of the kaehler metric is $SU(4)_I$, with isotropy group $SU(3)_I$. Consquently, the LCK metric Weyl connection's holonomy group for \MLc is also $SU(3)_I$. The isotropy representation of the group $SU(3)_I$ on the (real) tangent space is reducible to 
\[      
      \C^3 (V_1) + \C (V_0)
\]      
Thus, this is also the Weyl connection's holonomy reduction. (We are using the notation $V_1$ to refer to 
the complex defining representation of the group ($SU(3)$ in this case), $V_0$ to refer to the trivial representation, and the complex structure on $T\MLc$ to create an isomorphism between $Tp\MLc$ and $\C^4$.) 

At each point of \ML, $V_1$ and $V_0$ coincide with the distributions 
\begin{align*}
T &= ker(\theta+I\theta) \\
V &= \langle \theta^\#, I\theta^\# \rangle 
\end{align*}
where $T$ is mapped by the graph map onto $T\CP^3_{Io}$, and $b$ is kaehler on it. $V$ is the vertical 
distribution alone the graph map fibers. This holonomy reduction suggests that our LCK equations will 
take on a more transparent form when reduced to the distributions $T$ and $V$.

\subsubsection{LCHK Weyl Holonomy Reduction}\label{S:lchkholon}

As with the LCK case, the holonomy group of $(\MLc, h, \I,\J,\K)$ is defined by the isotropy subgroup of the isometery group of the hyperkaehler covering space $(\Hod, e, \I,\J,\K)$. The isometry group of the latter is $Sp(2)$, and thus the holonomy group is  the $Sp(1)_{iso}$ isotropy group. 

Like the distributions $T$ and $V$, associate with the toric fibration, that we used in the previous subsection, in the LCHK case, we can define distributions associated with the (negative) Swann fibration
\begin{align*}
D &= ker(\theta + \I\theta +\J\theta +\K\theta) \\
S &= <\theta^\#, \I\theta^\#, \J\theta^\#, \K\theta^\#>
\end{align*}
where $D$ is mapped by the Swann fibration onto $T\HP^1_o$, and $S$ is vertical to the
fibers of the negative Swann fibration.      

The isotropy group, $Sp(1)_{iso}$, acts trivially on the vertical distribution $S$, and via the $Sp(1)$ defining representation $V_1$ on $D$, so the holonomy reduction is
\[      
	\Hh (V_1) + \Hh (V_0).
\]
where the hypercomplex structure $(\I, \J, \K)$ has been used to create an isomorphism $Tp\ML \cong \Hh^2$. 

Again, this holonomy reduction suggests that our LCHK equations will take on a more transparent form 
when reduced to $D$ and $S$.
  
\subsection{Equations for Positive Sections}\label{S:pos}

We defined the maps $\PL$ using the non-zero parallel left spinors $\PsL$, sections of $\$^L$, which are also 
sections of $\$^-$, the negative Swann fibration of the hyperkaehler cover, $\Hod$, of $\ML$. We can 
also define similar maps $\PLp$ from non-zero parallel right spinors $\PsR$, derived from sections of the positive 
Swann fibration $\$^+$. These maps are also holomorphic (as we shall see), and thus harmonic, and 
solve the tension equation. We will show that the associated right and left Dirac equations are equivalent 
to the equations in physics that govern anti-fermion fields. 
      
The new maps fields differ from the old ones in several ways. Since $\PLp$ is a section of the positive 
Swann fibration, this also means that the hypercomplex structure is the positive one, derived from right quaternionic multiplication on the fiber.  This means that every complex structure is replaced by its negation, and composition of complex structure operators follows a different rule:
\[
\I \to -\I, \   \J \to -\J, \   \K \to -\K; \quad I \to -I; \quad (-\J)(-\I) = -\K
\]

However, we know that $-\I$ is also an element of $S^2 := a\I + b\J +c\K ; a^2+b^2+c^2 = 1$, so each negated 
complex structure is still part of the hypercomplex structure.
      
Changing $\$^- \to \$^+$ also changes the $P(\$^-)$ twistor space to the $P(\$^+)$ twistor space, and 
thus there is a parallel switch in the definition of the optical structure $O_\psi$, from a section of $
\mathbb{P}(\$^-)$ into a section of $\mathbb{P}(\$^+)$, so $O_\psi \to -O_\psi$. Our maps $\PLp$ 
remain holomorphic however, since $I  \in  S^2$ changes as well to $-I$. 

\subsubsection{Positive Sections and LCK holonomy}\label{S:poslck}
      
In addition, changing $Sp(1)^-$ to $Sp(1)^+$ changes the holonomy representations of both the LCK 
and LCHK Weyl connections. In the LCK case, changing the Swann fibration has an interesting effect on 
$SU(3)$ holonomy representations.  The holonomy representation of the connection preserves the
isomorphism  $T_x\ML \cong \CI^4$ induced by the complex structure. So changing the complex
structure to $-I$ changes the representation to the complex-conjugate defining representation:
\[      
      \C^3(V_1^*) + \C(V_0^*)
\]      
The resulting distributions are conjugate to $T$ and $V$, when $I$ is used to define the map of $T\ML \cong \C^4$, but on the real tangent bundle, they define the same distributions:
\begin{align*}
	T^* &= ker(\theta - I\theta)  = T\\
	V^* &= \langle \theta^\# + I\theta^\# \rangle = V
\end{align*}
When we discuss the holomophic tangent bundle and holonomy representations there, we will find that the conjugate distributions are in the anti-holomorphic bundle.

\subsubsection{Positive Sections and LCHK Holonomy}\label{S:poslchk}

In the LCHK case, changing from negative to positive vibrations has even less effect.  If we define new distributions, $D^+$ and $S^+$, we will find that they too are equal to the original $D$ and$S$ on the real tangent bundle:
\begin{align*}
D^+ &= ker(\theta - \I\theta - \J\theta - \K\theta) = D\\
S^+ &=  \langle \theta^\#, \Ip\theta^\#, \Jp\theta^\#, \Kp\theta^\# \rangle = S
\end{align*}
But, because for $Sp(1) \cong SU(2)$, the complex conjugate defining representation and the defining representation are not distinct, these two fibrations support the same holonomy representations: $\Hh(V_1) + \Hh(V_0)$.

So for both fibrations, the holonomy of the LCHK Weyl connection has $V_1$ holonomy on $D$ and 
trivial holonomy on $S$.  In the LCHK case, there are really only two types of $Sp(1)$ holonomy 
representations involved: $V_1$ and $V_0$, whereas in the LCK $SU(3)$ case, there are four: $V_1, 
V_1^*, V_0, V_0^*$.  Later, in section \ref{S:tens}, when we tensor LCK and LCHK isotropy 
representations together, we will find that there are only eight distinct types of tensor products, 
corresponding to quark doublets, quark singlets, anti-quark doublets, anti-quark singlets, lepton doublets, 
lepton singlets, anti-lepton doublets, and anti-lepton singlets.

\subsubsection{Positive Sections and Spinor Fields} \label{S:posspin}
There is, however, one additional effect of changing from positive to negative fibration, and that is on the parallel spinor field \Ps.  Since the change $\$^- \to \$^+$ is the same as $\$^L \to \$^R$, this change also changes $\Ps \to \Pst$.  

In the LCK case, this leads to the negative and positive tensor products: 
\begin{align*}
(V_1 + V_0) \xj \Ps &= V_1 \xj \PsL + V_1 \xj \PsR + V_0 \xj \PsR + V_0 \xj \PsR \\
(V_1^* + V_0^*) \xj \Pst &= V_1^* \xj \PsR + V_1^* \xj \PsL + V_1^* \xj \PsR + V_1^* \xj \PsL
\end{align*}
and the change from negative to positive sections, sends each component of the negative tensor product into the complementary one in the positive. Notice that all eight components are distinct.

The LCHK case is more interesting.  Since both negative and positive holonomy representations are the same, $(V_1 + V_0)$, there are only four distinct components in the tensor product:
\begin{align*}
(V_1 + V_0) \xj \Ps &= V_1 \xj \PsL + V_1 \xj \PsR + V_0 \xj \PsR + V_0 \xj \PsR \\
(V_1 + V_0) \xj \Pst &= V_1 \xj \PsR + V_1 \xj \PsL + V_1 \xj \PsR + V_1 \xj \PsL
\end{align*}
and switching from negative to positive sections doesn't change the total tensor product, although it does permute the components. 
       
So our Dirac equations in the LCK and LCHK cases for the positive maps $\PLp$ are
\begin{align*}      
	\gam \cdot \wkc{X^\mu}{(X^0 \xj \Pst)} &= 0 \\
	\gam \cdot \whk{X^\mu}{(X^0 \xj \Pst)} &= 0 
\end{align*}
where $X^0$ is the image of the time-like frame vector field under the map $d\PLp$, which is the same 
as the image under $d\PL$, since the chiral parallel spinor fields have the same representation as $\CI^2$ sections. 

$\wkc{}{}$ refers to the Weyl connection whose holonomy group, $SU(3)$, uses the complex-conjugate representation on $(T\ML, b, I)$.  We also introduce $\Pst$ to indicate that the spinor field changes when the map is positive.  This will be especially important in section \ref{S:tenseq}.

\section{Dirac-Higgs Equations on Complexified and Quaternified Tangent Bundles}\label{S:com}

Starting from the tension equation for (positive and negative) harmonic maps, we have derived Dirac equations, and Dirac-Weyl equations on the real tangent bundle of \ML, in both the LCK and LCHK cases, for both positive and negative sections.  We will now extend our analysis to holomorphic and anti-holomorphic bundles, in the LCK case, and to the ``hyperholomorphic'' tangent bundle in the LCHK case.

\subsection{Holomorphic and Anti-holomorphic Tangent Bundles of $\ML$}\label{S:comholo}

Clearly, simply complexifying the tangent bundle, $\C \otimes_\R T\ML$, does not affect the trace equation, but we would like to define equations for sections on the holomorphic and anti-holomorphic bundles $\Th$ and $\Tah$, because we know that $\wk{}{}$ preserves $I$. So the trace in the lifted tension equation vanishes separately on the holomorphic and anti-holomorphic tangent bundles. With this starting point, we can map our $T\ML$ vector fields derived from negative sections into the holomorphic bundle, $\Th$, using the map $1 - i I$, which is preserved by the Weyl connection, and replay the construction in subsection \ref{S:dirac}. In the LCK case, the resulting Dirac equation on the holomorphic tangent bundle is 
\[        
      \gam \cdot (\wk{\Zu}{\Zo \xj \Ps}) = 0        
\]      
where $\gam$ is still the lifted Clifford generator operating by Clifford multiplication of Dirac spinors, but the (tangent) Jordan bundle is now complexified. Now $\Zu$ is the canonical projection of $X^\mu$ onto the holomorphic tangent bundle, i.e., $\Zu = X^\mu - i I X^\mu$.  

The equation for positive sections naturally maps to the anti-holomorphic bundle, $\Tah$, since, as we saw in the previous section, right spinor fields as defined lead to a change in the definition of $I$ to $-I$. Thus, the map $1 - i I$ changes to $1 - i (-I) = 1 + i I$ and $Z \in \Th$ changes to $\bar{Z} \in \Tah$. So in the LCK case, the Dirac equation for positive sections is
\[        
      \gam \cdot (\wkc{\Zubar}{\Zobar \xj \Pst}) = 0        
\]      
Notice that since $X^\mu \in E$, $\Zu \in E := (1 - iI)E \subset \Th$, where we use $E$ for the distribution of the holomorphic tangent bundle. Similarly, $\Zubar \in \bar{E} := (1 + iI)E \subset \Tah$.
     
\subsection{Hyperholomorphic Tangent Bundle of $\ML$ } \label{S:comhypholo}

Likewise, ÒquaternifyingÓ the tangent bundle $\Hh \otimes_\R T\ML$ does not affect the trace equation, but finding the right expressions for equivalents of the holomorphic and anti-holomorphic tangent bundles is much trickier. Widdows' 2006 Oxford dissertation \cite{dw00}, supervised by Joyce, contains probably the best explication of the issues. The basic problem is that, while complexified bundles naturally split into two equal-dimensional bundles, each with the same real dimension as the original tangent space, quaternified tangent bundles split into two bundles, one of which has three times the number of dimensions as the other. Widdows calls these bundles $A$ and $B$. (Actually, he calls the cotangent equivalent $A$ and $B$.) At each point $x$, $A_x$ and $B_x$ are quaternionic vector spaces defined as follows:
\[      
      A_x : = \bigoplus_{J \in S^2} \Hh \otimes_{\C_j} T^{1,0}_{J x} \qquad     
      B_x : = \bigcap_{J \in S^2} \Hh \otimes_{\C_j} T^{0,1}_{J x}
\]      
We can also define complementary spaces $\bar{A}_x$ and $\bar{B}_x$ that also sum to $\Hh \otimes_\R T\ML$, but represent a different split into subbundles:
\[      
      \bar{A}_x : = \bigoplus_{J \in S^2} \Hh \otimes_{\C_j} T^{0,1}_{J x} \qquad     
      \bar{B}_x : = \bigcap_{J \in S^2} \Hh \otimes_{\C_j} T^{1,0}_{J x}
\]      
where in both case $T^{1,0}_{J x}$ and $T^{0,1}_{J x}$ are subspaces of $\Hh \otimes_\R T_x$ defined by the embedding 
\[
	\C \otimes_\R T_x \cong \C_j \otimes_\R T_x := (\R + j \R) \subset \Hh \otimes_\R T_x 
\]\[	 
	J = a\I + b\J + c\K \quad  j = a\bar{i} + b\bar{j} + c\bar{k} \qquad 1 = a^2+b^2+c^2 
\]
Clearly 
\[
J T^{1,0}_J = j T^{1,0}_J \quad \forall J, j 
\]
Note that the complex structures $J$ and $-J$ are both elements of the same hypercomplex structure defined by the two-sphere $S^2 = a\I + b\J +c\K ; a^2+b^2+c^2 = 1$. 

The bundles $A\ML$ and $\bar{A}\ML$ have six quaternionic dimensions, and are isomorphic to twelve-dimensional complex bundles. 

$A$ splits naturally into the sum of three two-quaternionic-dimensional bundles, $A_{\I}, A_{\J}, A_{\K}$, where at each point the subspaces can be defined by the mappings 
\[
      \alpha_n : T\ML \to A_n\ML;  \  \bar{\alpha}_n : T\ML \to \bar{A}_n\ML  \quad (n \in \{\I, \J, \K\})
\]\begin{align*}     
	\alpha_{\I} := 1 - \bar{i}\I, \quad \alpha_{\J} := 1 - \bar{j}\J, \quad \alpha_{\K} := 1 - \bar{k}\K \\
	\bar{\alpha}_{\I} := 1 + \bar{i}\I, \quad \bar{\alpha}_{\J} := 1 + \bar{j}\J, \quad \bar{\alpha}_{\K} := 1 + \bar{k}\K
\end{align*}
and in general
\[
 \alpha_J = (1 - j J), \bar{\alpha}_J = (1 + j J)\quad j = a\bar{i} + b\bar{j} + c\bar{k} \quad J = a\I + b\J + c\K 
\]     
     
 As was discussed in section \ref{S:pos}, in the LCHK case, the positive and negative sections of $\ML$ lead to the same $X^{\mu}$ vectors, and the positive complex structures are in the same $S^2$ as the negative complex structures, but with different sign. Thus the equivalent of the maps $1 - i I; 1 + i I$, which send $X^\mu \to \Zu ; \Zubar$ is, in the LCHK case, the two triplet of maps, $\alpha_n; \bar{\alpha}_n$:
\begin{align*}      
      \alpha_n: X^{\mu} \to (Q^{\mu}_{\I}, \ Q^{\mu}_{\J}, \ Q^{\mu}_{\K}) \\     
      \bar{\alpha}_n: X^{\mu} \to (\bar{Q}^{\mu}_{\I}, \ \bar{Q}^{\mu}_{\J}, \ \bar{Q}^{\mu}_{\K})      
\end{align*}      
            
So the hypercomplex form of our negative equations is actually a triplet of equations, one for each $A_n$, all with the same structure. The connection $\whk{}{}$ preserves the $A_n$ bundles, since it preserves the hypercomplex structure $\I, \J, \K$.  Thus, in the LCHK case, the Dirac equation for negative sections is:
\[      
      \gam \cdot (\whk{\Qu}{\Qo \xj \Ps}) = 0      
\]      
where $\Qo \in \{Q_{\I}^0, \ Q_{\J}^0,\  Q_{\K}^0\} \in A_n$. 

In the LCK holomorphic case, switching from $\$^-$ to $\$^+$ led us to exchange $I$ for $-I$, and thus to require equations for the holomorphic and anti-holomorphic bundles. In the LCHK case, this leads us to switch $\alpha_n \to \bar{\alpha}_n$, and thus $A_n \to \bar{A}_n$. So, when we apply our maps in the positive case we have the following triplet of Dirac equations for the positive sections:
 \[      
      \gam \cdot (\whk{\Qubar}{\Qobar \xj \Pst}) = 0.      
\]       
Notice that the connection, \whk{}{} does not change. This is consistent with the fact that the holonomy representation of $\whk{}{}$ is exactly the same in the positive and negative cases.

\subsection{Reduction of LCK Equations}\label{S:comholdist}

Our equations on $\Th$, $\Tah$ and $A$ are unsatisfactory as they sit, because neither Weyl connection preserves the associated metric, $b$ or $h$, so the holonomy representations are not unitary.  To compare our equations with physics, we would like to find connections that both preserve the holomorphic structure and the metric. In other words, we would like to find hermitian connections (see, \cite{pg97}, for example). 

We will examine the LCK case first. Our Dirac LCK equations for negative and positive sections are
\begin{align*}        
      \gam \cdot (\wk{\Zu}{\Zo \xj \Ps}) = 0  \\       
      \gam \cdot (\wkc{\Zubar}{\Zobar \xj \Pst}) = 0        
\end{align*}      
First, we will expand the Weyl connection in terms of the Levi-Civita connection (which preserves the metric), and then attempt to reduce the equations to distributions on which the induced metric is kaehler, and thus on which the Levi-Civita connection also preserves the complex structure. 

\subsubsection{Expanding the Weyl Connection} \label{S:expand}

Limiting ourself initially to negative sections, and thus to the holomorphic bundle, and using the definition of the Weyl connection from section \ref{S:harm}, we get
\[
 \gam \cdot (\lk{\Zu}{} -\half \theta_{\Zu})(\Zo \xj \Ps) - \half \gam \cdot ( \theta_{\Zo} \Zu \xj \Ps) 
 \]     
Let us examine the last term carefully. The Clifford generator, $\gam \cdot$ is acting by multiplication on the spinor component $\Ps$, but since the tensor product is defined over the Jordan algebra, we can re-write this term as 
\[
 \half (\theta_{\Zo} \Zu \xj (\gam \cdot \gamma^0 \cdot \Pst)) = \half \theta_{\Zo} (\Zu \cdot \sig)  \xj \Pst) = \theta_{\Zo} (\Zo \xj \Pst)
\]      
Notice, we have expanded $\Ps = \gamma^0 \cdot \Pst$, then moved the bivector $\gam \cdot \gamma^0$ across the tensor product to become $\sig$, multiplying vectors. Since $\Zu = \Zo \cdot \sig$ and $\sum_{\mu} \sig \cdot \sig = 2$, we are left with the final expression.

The other term involving the Lee form, $-\half \theta_{\Zu}$ is equivalent to $-\half iI\theta_{\Zu}$, where $iI\theta$ is the $U(1)$ lifted connection of the canonical bundle $K$ on $\CP^3_{Io}$ that we introduced in section \ref{S:lck}.

We can perform the same expansion for positive sections and the anti-holomorphic bundle, at which point our Dirac equations for positive and negative sections now have the form:
\begin{align*}
\gam \cdot (\lk{\Zu}{} - \half iI\theta_{Zu}) \Zo \xj \Ps &= \theta_{\Zo} \Zo \xj \Pst\\
\gam \cdot (\lkc{\Zubar}{} + \half iI\theta_{\Zubar}) \Zobar \xj \Pst &= \theta_{\Zobar} \Zobar \xj \Ps
\end{align*}

Notice that since $b$ is not kaehler, the Levi-Civita connections do not preserve the \Th and \Tah bundles inside the complexified tangent bundle (although the Weyl connection it is associated with does). We will remedy this problem in the next subsection.

\subsubsection{Reducing Connections to a Kaehler Distribution} \label{S:distrib}

We will employ the holomorphic and anti-holomorphic distributions of \Th and \Tah, defined by 
\begin{align*}
T &= ker (\theta + iI\theta) & \bar{T} &= ker(\theta- iI\theta) \\
V &= \langle \theta^\# - iI\theta^\# \rangle  & \bar{V} &= \langle \theta^\# + iI\theta^\# \rangle 
\end{align*}
to reduce our Dirac equations into derived equations operating independently on the distributions, based on connections that do preserve the holomorphic or anti-holomorphic character of the distributions.  

The important fact about $T$ and $\bar{T}$ is that, when reduced to them, the LCK metric $b$ is kaehler.  In addition, the distributions $V, \bar{V}$ are integrable, and their integral manifolds are just the torus fibers $\TLI$ of the original toric fibration of the graph map.  There is a natural isomorphism from $T$ to the holomorphic tangent bundle of $\CP^3_{Io}$ defined by the differential of the original graph map $Gr: \ML \to \CP^3_{Io}$. In addition, from our distribution $E$, associated with the map $\PL$ and the metric $g$, we can define horizontal projections $t, \bar{t}$ to $T, \bar{T}$, and vertical projections $v, \bar{v}$ to $V, \bar{V}$. (See Chapter 5 of Dragomir-Ornea for more details on distributions on LCK manifolds \cite{sd98}.)
      
Let us look at the properties of the $g$-trace of the Levi-Civita connection of $b$, $\lk{\Zu}{\Zu}$, a key element in the expansion of the original tension equation, and see how it reduces on the distributions $T$ and $V$. If we write the vector fields $\Zu$ as $\Zu = \Tu + \Vu$, where $\Tu$ and $\Vu$ are the horizontal and vertical projections of $\Zu$, we have:
\[       
	\lk{\Tu}{\Tu} + \lk{\Tu}{\Vu} + \lk{\Vu}{\Tu} + \lk{\Vu}{\Vu}
\]
The first term is isomorphic to the Levi-Civita connection of a kaehler metric on $\CP^3_{Io}$ acting on the holomorphic tangent bundle $\Th \CP^3$: 
\[
\lk{\Tu}{\Tu} = \lfs{\Tu}{\Tu} + \half \mathcal{V}[\Tu,\Tu] = \lfs{\Tu}{\Tu} 
\]
where ${\lfs{\Tu}{\Tu}}$ is the Levi-Civita connection of the kaehler metric on $T$ \cite[240]{ab87}. 

The second term is just $\partial_{\Tu} \Vu$, since $\lk{}{(\theta^\#-iI\theta^\#)} = 0$ and $\theta^\#$ has constant length \cite{sd98}.  However, it is isomorphic to a $U(1)$ connection, $\nabla^{U(1)}$, on the line bundle over $\CP^3_{Io}$ that is isomorphic to $V$, namely $\mathcal{O}(-2)$.

Now, in section \ref{S:sect}, when we constructed the equivalence class $|\PL|$ of holomorphic sections, we composed with the toric group $\TLI$, so the toric action preserves $\Zu$. Thus the vertical component $\Vu$ is Killing and the horizontal component $\Tu$ has vanishing Lie derivative with respect to $\theta^\#-iI\theta^\#$. Therefore, the third term in our expression vanishes, since 
\[
      \lk{\Vu}{\Tu} = \lk{\Tu}{\Vu} + [\Vu, \Tu] = d_{\Tu} \Vu - d_{\Tu} \Vu
\]      
as does the fourth term, since $\lk{\Vu}{\Vu} = -\frac{1}{2} [\Vu, \Vu] = 0$. We are left with just 
\[
\lk{\Zu}{\Zu} = \lk{\Tu}{\Tu} + \partial_{\Tu} \Vu, 
\]
which looks like two connection forms on $\CP^3_{Io}$, acting on sections of different two different bundles. The first is acting on sections of the holomorphic tangent bundle, $\Th\CP^3_{Io}$; the second on sections of the the line bundle $\mathcal{O}(-2)$.

So let us use the horizontal projection $t$ of $\Zu$ to write $\lk{}{}$ as the sum of two connections that are pulled back from $T$ along the horizontal projection: 
\begin{align*}
\lk{\Zu}{\Zu} &= t^* \lfs{\Zu}{\Tu} + t^* \nabla^{U(1)}_{\Zu} \Vu \\
				&:= \lfs{(h\Zu)}{\Tu} + \nabla^{U(1)}_{(h\Zu)} \Vu
\end{align*}

Clearly $t^* \lfs{}{}$ as defined preserves $k$, the kaehler metric on $\Th\CP^3_{Io}$, as well as $I$ acting in $T$, and $\Th \CP^3_o$. Similarly, $t^* \nabla^{U(1)}$ preserves $I$ acting in $V$, which as we saw in section \ref{S:lchk}  is isomorphic to $\mathcal{O}(-2)$ on $CP^3_{Io}$.  We will use this new representation of the Levi-Civita connection to re-write our Dirac equations into a more familiar form. We get two independent equations for negative sections and two for positive sections:
\begin{align*}      
      \gam \cdot (t^*\lfs{\Zu}{} -\half iI\theta_{\Zu} \To \xj \Ps) &= \theta_{\Zo}(\To \xj \Pst) \\     
      \gam \cdot (t^*\nabla^{U(1)}_{\Zu} -\half iI\theta_{\Zu} \Vo \xj \Ps) &= \theta_{\Zo}(\Vo \xj \Pst) \\
      \gam \cdot (t^*\lfsc{\Zubar}{} + \half iI\theta_{\Zubar} \Tobar \xj \Pst) &= \theta_{\Zobar}(\To \xj \Ps)\\      
      \gam \cdot (t^*\bar{\nabla}^{U(1)}_{\Zubar} + \half iI\theta_{\Zubar} \Vobar \xj \Pst) &= \theta_{\Zobar}(\Vo \xj \Ps)
\end{align*}            
where we have used the same symbol, $t$, for the horizontal maps $E \to T$ and $\bar{E} \to \bar{T}$. Notice that, since $\Ps$ and $\Pst$ are composed of $\PsL$ and $\PsR$, these four Dirac-Higgs equations are equivalent to eight coupled Dirac-Weyl-Higgs equations (which I will refrain from writing down), and that all eight are distinct.

\subsubsection{A Comparison with Physics} \label{S:physicslck}

At this point it is convenient to compare our Dirac equations with the  form they take in physics. Our equations are defined on distributions $E$ and $\bar{E}$ of the complex tangent spaces $\Th \ML$ and $\Tah \ML$.  They have been derived from a lifted tension equation of the form $\wk{}{(g^*)} - \lcl{}{(g^*)} = 0$, where the connections and the cometic, $g^*$ are both defined on \ML. The equations of physics, on the other hand, describe sections of arbitrary (non-tangent) complex vector bundles over real four-dimensional space-time manifolds. 

We can, however, pull back all connections and forms from $\ML$ to the original Lorentzian base manifold $\HP^1_o$, along the maps $\rho := (1 - iI) \circ \PL$ and $\bar{\rho} := (1 + iI) \circ \PLp$. The connections now are acting on pullback bundles. The result is
\begin{align*}      
      \gam \cdot (t^*\lfs{e^\mu}{} -\half iI\theta_{e^\mu}) (\To \xj \Ps) &= \theta_{\Zo}(\To \xj \Pst)\\      
      \gam \cdot (t^*\nabla^{U(1)}_{e^\mu} -\half iI\theta_{e^\mu}) (\Vo \xj \Ps) &= \theta_{\Zo}(\Vo \xj \Pst)\\
      \gam \cdot (t^*\lfsc{e^\mu}{} + \half iI\theta_{e^\mu}) (\Tobar \xj \Pst) &= \theta_{\Zobar}(\To \xj \Ps)\\      
      \gam \cdot (t^*\bar{\nabla}^{U(1)}_{e^\mu} + \half iI\theta_{e^\mu}) (\Vobar \xj \Pst) &= \theta_{\Zobar}(\Vo \xj \Ps)
\end{align*}            

\begin{notation}
We have not changed the symbols for the connections and forms or fields, even though the connections and forms now represent pulled back versions, via $\rho^*$ and $\bar{\rho}^*$, and $\Zo \in \rho^* \Th$, $\Zobar \in \bar{\rho}^* \Tah$ and $\Ps, \Pst \in \$$ are now sections of pulled back bundles. This follows our general policy of usually not distinguishing between the original and lifted forms of other structures.
\end{notation}

The pullback connections are complex Lie Algebra-valued forms on $\HP^1_o$ that act on sections of the pull-back bundles. The forms are now complex-valued, so they act on the complex pull-back bundles by complex multiplication. By these pullbacks we are treating $\Th$ and $\Tah$  simply as complex vector bundles over $\HP^1_o$. 

Let us define Dirac operators as follows, using the Dirac ``slash'' notation:
\begin{align*}
\Db &:= \gam \cdot (t^*\lfs{}{})_{e^\mu}	 \\
\D &:= \gam \cdot (t^*\nabla^{U(1)})_{e^\mu}	 \\
\Dbc &:= \gam \cdot (t^*\lfsc{}{})_{e^\mu}	 \\
\Dc &:= \gam \cdot (t^*\bar{\nabla}^{U(1)})_{e^\mu}
\end{align*}
Let us also define the ``dotted'' theta connection, equivalent to ``slashed''. (We use dotted, because slashed is very hard to read.)
\[
\half iI\dtc := \half iI \gam \cdot \theta_{X^\mu}
\]

Then, the Dirac operator form of the pulled back equations is:
\begin{align*}      
      (\Db -\half iI\dtc) (\To \xj \Ps) &= \theta_{\Zo}(\To \xj \Pst) \\     
      (\D -\half iI\dtc) (\Vo \xj \Ps) &= \theta_{\Zo}(\Vo \xj \Pst)\\
      (\Dbc + \half iI\dtc) (\Tobar \xj \Pst) &= \theta_{\Zobar}(\Tobar \xj \Ps)\\      
      (\Dc + \half iI\dtc) (\Vobar \xj \Pst) &= \theta_{\Zobar}(\Vobar \xj \Ps)
\end{align*}

The pulled-back form of the equations is certainly familiar from physics.  We have vector bundles over a four-dimensional Lorentzian manifold acted upon by Dirac operators associated with an $SU(3)$-connection. The bundles hold irreducible unitary representations of $SU(3)$, which thus acts as a ``gauge group.'' We will thus identify the pulled back $T$ bundle, containing $SU(3)$ triplets with \emph{quarks}, $\bar{T}$ with \emph{anti-quarks}. The pulled back $V$ bundle, containing $SU(3)$ singlets, we will identify with \emph{leptons}; $\bar{V}$ with \emph{anti-leptons}. (All the pulled back forms are complex valued since they still take values in in the complex pulled back bundles.) 

In each case, we have a Dirac operator for a gauge connection. For quarks and anti-quarks the connection tensors an $U(3) = SU(3)U(1)$-holonomy connection, $t^*\lfs{}{}$, pulled back from $T \cong \Th \CP^3_o$, with another connection, $-\half iI\theta$, associated with the $U(1)$ gauge group of the $\mathcal{O}(-2)$ line bundle on $\CP^3_{Io}$. For leptons we just have the $U(1)$ gauge group connection $t^*\nabla^{U(1)}$ tensored with the $U(1)$ theta connection, $-\half iI\theta$.

Because we will often be comparing our equations with physics, it is convenient to think of the symbol $\Db -\half iI\dtc$ as standing for the Dirac operator of an $SU(3)$ connection tensored with a $U(1)U(1)$ connection. We will identify the remnant $SU(3)$-gauge connection as the ÒcolorÓ interaction, while the $U(1)U(1)$ product connection $\nabla^{U(1)} \otimes -\half iI\theta$ we will call the ÒhypercolorÓ interaction. 

All fields are Dirac-spinors. The equations couple $\Ps$ spinors with $\Pst$ spinors. As is well-known in physics, each such equation is equivalent to two coupled equations: left-spinors, \PsL, coupled to right-spinors, \PsR. Thus in terms of left and right spinors, we have eight equations: left quarks are coupled to right quarks, left anti-quarks with right anti-quarks, left leptons with right leptons, left anti-leptons with right anti-leptons. 

The coupling field, $\theta_{\Zo}, \theta_{\Zobar}$, we will identify with part of the the Higgs field. Later, we will show how our hypercolor and Higgs coupling are part of the Standard Model.

\subsection{Reduction of LCHK Equations to Distributions}\label{S:comhypdist}
Let us turn now to our triplets of Dirac equations in the LCHK case
\begin{align*}      
      \gam \cdot (\whk{\Qu}{\Qo \xj \Ps}) &= 0  \\    
      \gam \cdot (\whk{\Qubar}{\Qobar \xj \Pst}) &= 0      
\end{align*}      
where at any point $\Qu \in A_n\ML; \Qubar \in \bar{A}_n\ML$.

We would like to perform the same reduction on these equations that we did in the LCK case, namely, to expand them in terms of the Levi-Civita connection, and then use distributions in in $A_n$ with special properties to create independent equations that preserve the distributions. In the LCK case we used to good effect the kaehler properties of the distribution $T$ along with the Killing properties of vector fields in the vertical distribution $V$. We shall find similar distributions in the LCHK case.

\subsubsection{Expanding the Weyl Connection}\label{S:expandlchk}
 
After working through the same steps that we did in section \ref{S:expand}, and bearing in mind that there is no distinction in the LCHK case between the connections for positive and negative sections, we arrive at the following expanded Dirac-Higgs LCHK equations for negative and positive sections:
\begin{align*}
\gam \cdot (\lhk{\Qu}{} - \half i_n \I_n \theta_{Qu}) \Qo \xj \Ps = \theta_{\Qo} \Qo \xj \Pst\\
\gam \cdot (\lhk{\Qu}{} - \half i_n \I_n \theta_{Qu}) \Qo \xj \Pst = \theta_{\Qo} \Qo \xj \Ps\\
\gam \cdot (\lhk{\Qubar}{} - \half i_n \I_n \theta_{Qu}) \Qobar \xj \Pst = \theta_{\Qo} \Qo \xj \Ps\\
\gam \cdot (\lhk{\Qubar}{} - \half i_n \I_n \theta_{Qu}) \Qobar \xj \Ps = \theta_{\Qo} \Qo \xj \Pst
\end{align*}
where $i_n \in \{\bar{i}, \bar{j} \bar{k}\}$ and $\I_n \in \{\I, \J, \K\}$ and the forms take quaternion values when evaluated on vector fields.

\subsubsection{Reducing Connections to Quaternionic Distributions}\label{S:lchkdistrib}

We now want to follow the same procedure that we did in section \ref{S:distrib}, for the LCK case, this time using distributions in $A_n$ with special quaternionic properties, just as $T$ and $V$ had special properties in $\Th$.

In section \ref{S:lchkholon} we introduced real distributions $D$ and $S$ on $T\ML$ that were tied to the holonomy of the LCHK Weyl connection.
\begin{align*}
D &= ker(\theta + \I\theta + \J\theta + \K\theta) \\
S &= <\theta^\#, \I\theta^\#, \J\theta^\#, \K\theta^\#>
\end{align*}

There are some differences between $D$ and the $T$ of the LCK case. Most importantly, the reduction of the metric $h$ to $D$ is a quaternionic kaehler metric $q$, not a hyperkaehler metric. This means that the Levi-Civita connection preserves the whole $S^2$ sphere of complex structures in $End(D)$, but not any one complex structure. (In physics terms, the distinction between ``upness'' and ``downness'' is not preserved, although the distinction between ``doubletness'' and ``singleness'' is. We will see later, that selecting a particular complex structure $I$, leads to the preservation of a particular ``up'' and ``down'' distinction.)

These distributions can be mapped to the hyperholomorphic bundles $A_n, \bar{A}_n$ easily, by the definitions:
\[      
      D_n := \alpha_n(D) \quad  S_n := \alpha_n(S) \quad \bar{D}_n := \bar{\alpha}_n(D)\quad \bar{S}_n := \bar{\alpha}_n(S)
\]      
We can follow our LCK procedure and split 
\[
\Qu = \Du + \Su \quad \Qubar = \Dubar + \Subar
\] 
Then, after evaluating \lhk{}{} on these distributions, we can write it, as we did in the LCK case, as the sum of two connections:
\[
\lhk{\Qu}{\Qu} = du^* \lcq{\Qu}{\Du} + du^* \nabla^{Sp(1)}_{\Qu} \Su \quad \lcq{\Qubar}{\Qubar} = du^* \lcq{\Qubar}{\Dubar} + du^* \nabla^{Sp(1)}_{\Qubar} \Subar 
\]
where $\lcq{}{}$ is the Levi-Civita connection of the quaternionic-kaehler metric $q$ on $\HP^1_o$, and $\nabla^{Sp(1)}$ is the $Sp(1)$ connection on the $\Hh$-line bundle on $\HP^1_o$ isomorphic to $S_n$, which is, of course, the tautological $\Hh$ bundle. We have uses $du$ for the horizontal projection to both $D$ and $\bar{D}$. 

Based on this representation of \lhk{}{}, we can rewrite the Dirac-Higgs equations in the LCHK case for positive and negative sections as:
\begin{align*}      
      \gam \cdot (du^*\lcq{\Qu}{} -\half i_n \I_n\theta_{\Qu} \Do \xj \Ps) &= \theta_{\Qo}(\Do \xj \Pst) \\     
      \gam \cdot (du^*\nabla^{Sp(1)}_{\Qu} -\half i_n \I_n\theta_{\Qu} \So \xj \Ps) &= \theta_{\Qo}(\So \xj \Pst) \\
      \gam \cdot (du^*\lcq{\Qubar}{} + \half i_n \I_n\theta_{\Qubar} \Dobar \xj \Pst) &= \theta_{\Qobar}(\Do \xj \Ps)\\      
      \gam \cdot (du^*\nabla^{Sp(1)}_{\Qubar} + \half i_n \I_n\theta_{\Qubar} \Sobar \xj \Pst) &= \theta_{\Qobar}(\So \xj \Ps)
\end{align*}

\subsubsection{A Comparison with Physics}\label{S:physicslchk}

We can also compare these equations with physics, by pulling the four distinct LCHK equations back to the Lorentzian manifold $(\HP^1_o, g)$, as we did in the LCK case. This time, the map we are pulling back along is $\kappa_n := (1 - i_n \I_n) \circ \PL$, and the resulting equations are:
\begin{align*}      
      \gam \cdot (du^*\lcq{e^\mu}{} -\half i_n \I_n\theta_{e^\mu} \Do \xj \Ps) &= \theta_{\Qo}(\Do \xj \Pst) \\     
      \gam \cdot (du^*\nabla^{Sp(1)}_{e^\mu} -\half i_n \I_n\theta_{e^\mu} \So \xj \Ps) &= \theta_{\Qo}(\So \xj \Pst) \\
      \gam \cdot (du^*\lcq{e^\mu}{} + \half i_n \I_n\theta_{e^\mu} \Dobar \xj \Pst) &= \theta_{\Qobar}(\Do \xj \Ps)\\      
      \gam \cdot (du^*\nabla^{Sp(1)}_{e^\mu} + \half i_n \I_n\theta_{e^\mu} \Sobar \xj \Pst) &= \theta_{\Qobar}(\So \xj \Ps)
\end{align*}            


Using the Dirac operator slash notation from the previous section,, with the following addition:
\[
\dd := \gam \cdot (t^*\nabla^{Sp(1)})_{e^\mu}	 
\]
we can write the LCHK equivalent equations as:
\begin{align*}      
      (\Dh -\half i_n \I_n\dth) (\Do \xj \Ps) &= \theta_{\Qo}(\To \xj \Pst) \\     
      (\dd -\half i_n \I_n\dth) (\So \xj \Ps) &= \theta_{\Qo}(\Vo \xj \Pst)\\
      (\Dh + \half i_n \I_n\dth) (\Dobar \xj \Pst) &= \theta_{\Qobar}(\Dobar \xj \Ps)\\      
      (\dd + \half i_n \I_n\dth) (\Sobar \xj \Pst) &= \theta_{\Qobar}(\Sobar \xj \Ps)
\end{align*}

The pulled-back form of these equations are also familiar from physics.  The bundles hold irreducible representations of $Sp(1)$, which is the gauge group. We identify the pulled-back $D_n$ bundles, containing the $Sp(1)_{iso}$ non-trivial representations, with \emph{weak doublets} of all three generations. The pulled back $S_n$ bundles, containing $Sp(1)_{iso}$ trivial representations, we will identify with \emph{weak singlets}. The $\bar{D}_n$ and $\bar{S}_n$ bundles represent anti-fermions.

The bundles $D_n$ are acted upon by the Levi-Civita connection, $\lcq{}{}$, of the quaternionic kaehler metric $q$ and by the quaternionic theta connection $i_n I_n \dth$, while the bundles $S_n$ are acted upon by the tautological $\Hh$-line bundle connection, $\dd{}{}$ and the  theta connection. 

In each case, we have a Dirac operator for a gauge connection. For weak doublets the connection combines an $Sp(1)_{iso}Sp(1)$-holonomy connection $t^*\lhk{}{}$, pulled back from $D_n \cong T \HP^1_o$, with the theta connection, $-\half i_n \I_n \theta$, associated with the $Sp(1)$ gauge group of the tautological $\Hh$-line bundle isomorphic with $S$. For weak singlets, we just have the $Sp(1)$ gauge group connection $t^*\nabla^{Sp(1)}$ tensored with the $Sp(1)$ theta connection.

To compare with physics, it is convenient to think of the symbol $\dd -\half i_n \I_n\dth$ as standing for the Dirac operator of an $Sp(1)_{iso}$ connection tensored with an $Sp(1)Sp(1)$ connection. We will identify the remnant $Sp(1)_{iso}$-gauge connection as the ÒweakÓ interaction, while the $Sp(1)Sp(1)$ product connection $\nabla^{Sp(1)} \otimes -\half i_n I_n \theta$ we will call the ÒhyperspinÓ interaction.

These equations also couple $\Ps$ spinors with $\Pst$ spinors.The coupling field, $\theta_{\Qo}, \theta_{\Qobar}$, we will again identify with part of the the Higgs field. Later, we will show how our hyperspin and Higgs coupling match the Standard Model.

\section{Tensor Product Bundles and Equations} \label{S:tens}

Through this point in our analysis of \ML and its structures, we have treated the complex and hypercomplex structures relatively independently, although we have made use from time to time of important links between them (for example, between the choice of Swann fibration and the switch $I \to -I$). Motivated by the striking similarities between our transformed tension equations and the Dirac-Higgs equation of physics, we will now introduce a construction that combines the LCK and LCHK structures into one. 
      
Our analysis has shown that the holonomy group $SU(3)$ of the LCK Weyl connection $\wk{}{}$ has a role in our equations that is very similar to the role of the color interaction gauge group of the classical field equations of the Standard Model. Similarly, the holonomy group $Sp(1)$ of the LCHK Weyl connection $\whk{}{}$ has role similar to that of the weak-isospin gauge group. In physics, the combined action of color and weak-isospin is by a combined connection for a product gauge group $SU(3)Sp(1)$ acting on fields with different indices for their color and weak-isospin components. In other words, these classical fields are represented as tensors, sections of the tensor product of a color representation bundle and a weak isospin representation bundle. In fact, as we shall see in this section, these fields look exactly like sections of the tensor products $\Th \otimes_{\C} A_n$ (three generations of fermions) and $\Tah \otimes_\C A_n$ (three generations of anti-fermions).
      
From a mathematical perspective, taking such a tensor product seems to be a little unusual, but it has the advantage of displaying the holomorphicity properties of the complex structure (which led to the harmonic map in the first place) and the quaternionic character of the hypercomplex structure in one bundle.

\subsection{Tensor Product Bundles and Standard Model Fermions}
Let us closely examine the bundles we will be tensoring. 

Throughout this paper, we have re-written the tension equation of our harmonic map as a Dirac equation or Dirac-Higgs equation, by tensoring $\$ = \$^L + \$^R$ spinor bundles to the tangent bundles. We will now do that with the above tensor product tangent bundles. Our equations will then involve tensor-products of the connections we used earlier, which were themselves tensor product connections of LCK or LCHK Weyl connections with the Lorentz spin connection. The somewhat tricky part is defining the tensor \emph{triple product} bundles so that the associated connections behave ``properly,'' that is, so that the associated Dirac-Higgs equations reduce to the equations we have already derived on the \emph{double product} bundles.

Let us consider our sets of negative and positive LCK Dirac Higgs equations. The connections and Dirac operators are defined on double product bundles, $(T + V) \xj \$$ and $(\bar{T} + \bar{V}) \xj \$$:
\begin{align*}      
      (\Db -\half iI\dtc) (\To \xj \Ps) &= \theta_{\Zo}(\To \xj \Pst) \\     
      (\D -\half iI\dtc) (\Vo \xj \Ps) &= \theta_{\Zo}(\Vo \xj \Pst)\\
      (\Dbc + \half iI\dtc) (\To \xj \Pst) &= \theta_{\Zobar}(\To \xj \Ps)\\      
      (\Dc + \half iI\dtc) (\Vobar \xj \Pst) &= \theta_{\Zobar}(\Vo \xj \Ps)
\end{align*}
On the left-hand side of the equations, the $SU(3)$ triplet field, $\To$ is tensored with $\Ps = \PsL + \PsR$ in the negative section and $\Pst = \PsR + \PsL$, in the positive section, as is the singlet field, $\Vo$. 

Now consider the sets of negative and positive LCHK equations. The connections and Dirac operators are defined on the double product bundles, $(D_n + S_n) \xj \$$ and $(\bar{D}_n + \bar{S}_n) \xj \$$:
\begin{align*}      
      (\Dh -\half i_n \I_n\dth) (\Do \xj \Ps) &= \theta_{\Qo}(\To \xj \Pst) \\     
      (\dd -\half i_n \I_n\dth) (\So \xj \Ps) &= \theta_{\Qo}(\Vo \xj \Pst)\\
      (\Dh + \half i_n \I_n\dth) (\Dobar \xj \Pst) &= \theta_{\Qobar}(\Dobar \xj \Ps)\\      
      (\dd + \half i_n \I_n\dth) (\Sobar \xj \Pst) &= \theta_{\Qobar}(\Sobar \xj \Ps)
\end{align*}

We need to define triple product bundles in such a way that their Dirac-Higgs equations for sections reduce to these equations on sections of these double product product bundles. We will take tensor products over the complex numbers $\C_i \subset \Hh; i = a\bar{i} + b\bar{j} + c\bar{k}$.

Let us begin with the obvious products:
\begin{align*}
(T + V) \xc (D_n + S_n) \xj \Ps = (T + V) \xc (D_n \xj \PsL + S_n \xj \PsL + D_n \xj \PsR + S_n \xj \PsR) \\
(\bar{T} + \bar{V}) \xc (\bar{D}_n + \bar{S}_n) \xj \Pst = (\bar{T} + \bar{V}) \xc (\bar{D}_n \xj \PsR + \bar{S}_n \xj \PsR + \bar{D}_n \xj \PsL + \bar{S}_n \xj \PsL)
\end{align*}

Our equations contain sections from all sixteen of these triple product half-spinor bundles. However, by taking a clue from physics, we can halve the number of triple product bundles we have to deal with, and cut the number of bundles down to eight. 

Our equations, in both the LCK and LCHK case, couple $\$^L$ fields with $\$^R$ fields in the same vector bundle. That is, $T \xj \$^L$ with $T \xj \$^R$, $\bar{D} \xj \$^L$ with $\bar{D} \xj \$^R$, and so on. Since $D_n$ and $S_n$ are both one-dimensional quaternionic sub-bundles of $Q_n$, there is a straightforward way to change the coupling of $D_n$ with $D_n$ to a coupling of $D_n$ with $S_n$. If our equations are rewritten in this way, we only need one copy each of $D_n$ and $S_n$ in our tensor products, in order to include all the sections in our equations, since $D_n \xj \$^L$ will couple to $S_n \xj \$^R$ instead of $D_n \xj \$^R$, and so on.  We only need the following tensor products:
\begin{align*}
(T + V) \xc (D_n + S_n) \xj \Ps = (T + V) \xc (D_n \xj \PsL + S_n \xj \PsR) \\
(\bar{T} + \bar{V}) \xc (\bar{D}_n + \bar{S}_n) \xj \Pst = (\bar{T} + \bar{V}) \xc (\bar{D}_n \xj \PsR + \bar{S}_n \xj \PsL)
\end{align*}
In the section \ref{S:higgs}, we will cover how to rewrite our Dirac-Higgs equations in such a manner that they couple $D_n$ to $S_n$.

We can identify sections of these tensor product bundles with the fermion fields of physics:
\begin{align*}
	T \xc (D_n \xj \$^L + S_n \xj \$^R)	& &q_n^L, q_n^R	& &\text{left and right quarks}\\
	V \xc (D_n \xj \$^L + S_n \xj \$^R) & &l_n^L, l_n^R &	&\text{left and right leptons}\\
	\bar{T} \xc (\bar{S}_n \xj \$^L + \bar{D}_n \xj \$^R) &	&\bar{q}_n^L, \bar{q}_n^R &	&\text{left and right anti-quarks}\\
	\bar{V} \xc (\bar{S}_n \xj \$^L + \bar{D}_n \xj \$^R) &	&\bar{l}_n^L, \bar{l}_n^R &	&\text{left and right anti-leptons}
\end{align*}

\subsection{Tensor Product Dirac-Higgs Equations}\label{S:tenseq}

By assembling the following tensor product fields in these bundles, we can derive Dirac-Higgs equations for the classical fermion fields of physics that will reduce to our existing equations.
\begin{align*}
q_n^L &:= \To \xc \Do \xj \PsL  & q_n^R &:= \To \xc \So \xj \PsR  \\
 l_n^L &:= \Vo \xc \Do \xj \PsL & l_n^R &:= \Vo \xc \So \xj \PsR\\
\bar{q}_n^L &:= \Tobar \xc \Sobar \xj \PsL & \bar{q}_n^R &:= \Tobar \xc \Dobar \xj \PsR \\
\bar{l}_n^L &:= \Vobar \xc \Sobar \xj \PsL & \bar{l}_n^R &:= \Vobar \xc \Dobar \xj \PsR 
\end{align*}

Next, we define the tensor product Dirac-Weyl operators that are associated with these tensor product bundles. First, the Dirac-Weyl operators for quarks and anti-quarks:
\begin{align*}
\DbhL &:= \sig \cdot [(t^*\lfs{}{}) \xc (du^* \lcq{}{})]_{X^\mu}	 \\
\DbR &:= \sigr \cdot [(t^*\lfs{}{}) \xc (du^* \nabla^{Sp(1)})]_{X^\mu}	\\
\DbLc &:= \sig \cdot [(t^*\lfsc{}{}) \xc (du^* \nabla^{Sp(1)})]_{X^\mu}	\\
\DbhRc &:= \sigr \cdot [(t^*\lfsc{}{}) \xc (du^* \lcq{}{})]_{X^\mu} 
\end{align*}
 Then, the Dirac-Weyl operators leptons and anti-leptons:
\begin{align*}
\DhL &:= \sig \cdot [(t^*\nabla^{U(1)}) \xc (du^* \lcq{}{})]_{X^\mu}	 \\
\DR &:= \sigr \cdot [(t^*\nabla^{U1)}) \xc (du^* \nabla^{Sp(1)})]_{X^\mu}	\\
\DLc &:= \sig \cdot [(t^*\bar{\nabla}^{U(1)}) \xc (du^* \nabla^{Sp(1)})]_{X^\mu}	\\
\DhRc &:= \sigr \cdot [(t^*\bar{\nabla}^{U(1)}) \xc (du^* \lcq{}{})]_{X^\mu} 
\end{align*}
We will also use the ``dotted` theta connections that we defined earlier, which again correspond to ``slashed'', but are somewhat easier to read.
\begin{align*}
\half iI\dtcl &:= \half iI \sig \cdot \theta_{X^\mu}\\
\half i_n I_n \dthl &:= \half i_n I_n \sig \cdot \theta_{X^\mu}\\
\half iI\dtcr &:= \half iI \sigr \cdot \theta_{X^\mu}\\
\half i_n I_n \dthr &:= \half i_n I_n \sigr \cdot \theta_{X^\mu}
\end{align*}

Let us now examine the final pulled-back Dirac-Higgs equations that result for each of these four types of tensor-product fermions, after we expand the tensor product Dirac operators into a form with the covariant operator associated with the connection written out in full:

\textbf{Quarks and Leptons}:
\begin{align*}      
(\DbhL - \half iI\dtcl - \half i_n \I_n\dthl) q_n^L &= (\Higgs) (\To \xc \Do \xj \PsR) \\
(\DbR - \half iI\dtcr - \half i_n \I_n\dthr) q_n^R &= (\Higgs) (\To \xc \So \xj \PsL) \\
(\DhL - \half iI\dtcl - \half i_n \I_n\dthl) l_n^L &= (\Higgs) (\Vo \xc \Do \xj \PsR) \\
(\DR - \half iI\dtcr - \half i_n \I_n\dthr) l_n^R &= (\Higgs) (\Vo \xc \So \xj \PsL)
\end{align*}

\textbf{Anti-quarks and Anti-leptons}:
\begin{align*}      
(\DbLc + \half iI\dtcl + \half i_n \I_n\dthl) \bar{q}_n^L &= (\Higgs) (\Tobar \xc \So \xj \PsR) \\
(\DbhRc + \half iI\dtcr + \half i_n \I_n\dthr) \bar{q}_n^R &= (\Higgs) (\Tobar \xc \Do \xj \PsL) \\
(\DLc + \half iI\dtcl + \half i_n \I_n\dthl) \bar{l}_n^L &= (\Higgs) (\Vobar \xc \So \xj \PsR) \\
(\DhRc + \half iI\dtcr + \half i_n \I_n\dthr) \bar{l}_n^R &= (\Higgs) (\Vobar \xc \Do \xj \PsL)
\end{align*}
Notice that in this form, the $D_n \xj \$^L$ sections are still coupled to $D_n \xj \$^R$ sections, via a Higgs field that is the product of the complex scalar function $\theta^{\C}_{X^0}$ with quaternionic function $\theta^{\Hh}_{X^0}$. We will remedy that in the next section.  

The Higgs fields of physics is represented here by the product of the Higgs field ``connections'', $\theta^{\C}_{\Zo}$ and $\theta^{\Hh}_{\Qo}$.

\subsubsection{The Higgs ``Field''}\label{S:higgs}

As we discussed in the previous subsection, by only using half as many bundles, we give up immediate coupling of equations. The vector-valued spinor fields on the right sides of the equations are not in the tensor product bundles, so they are not in our collection of fermion fields. (The basic issue is that on the right side, the $\Dio$ and $\Sio$ components of the tensor product fields would have to be switched for the product field to be in one of our product bundles.) To accomplish this, we need to digress a little to discuss the way the Higgs field enters the Standard Model.

In the physics literature, the pullbacks of $S_I$ and $D_I$ are not conceived of as subbundles of a single bundle, but rather as independent bundles, where $S_I$ is trivial. Sections of the two trivial complex one-dimensional subbundles of $S_I$ are conceived of simply as scalar complex-valued functions. Under this perspective, the only true Òvector bundleÓ is the pullback of $D_n$, and the only vector-valued field is $\Do$. Consequently, in the physics literature, the switch from $\Do$ to $\So$ is seen as turning a vector field into a pair of scalar fields, by taking inner products of $\Do$ with a ÒHiggs doubletÓ field. The reverse process of switching from $\So$ to $\Do$ is seen as multiplying complex scalar fields by the same Higgs doublet field.  So the end result is that our first equation is interpreted as that the $\DbhL$ connection turns a $D_n$ vector field into another $D_n$ vector field, namely, the Higgs field, multiplied by a scalar field (the section of $S_n$). The expressions to accomplish this task are quite involved [NOTE].

A simpler approach, which we will adopt here, is to define the ÒswitchÓ operator $Sw$. It sends a quaternionic $D_n$ one-vector into the quaternionic $S_n$ one-vector with the same quaternionic magnitude. In terms of the local basis $(\So / || \So ||, \Do / || \Do ||)$ at a point p of \ML, $Sw$ has the representation $\begin{pmatrix} 0 & 1\\ 1 &  0 \end{pmatrix}$. Clearly $Sw^{-1} = Sw$. Thus:
\begin{align*}      
      \Do &= (q_{n d})(q_{n s}s)^{-1} Sw \So  \\   
      \So &= (q_{n s})(q_{n d})^{-1} Sw \Do
\end{align*}      
where the $q_{n d}, q_{n s}$ quaternion-valued functions are the quaternionic magnitudes of the $\Do$ and $\So$ fields at each point. Since our Higgs field is just another quaternion-valued function itself, it can be multiplied with the re-scaling factors.  Thus:
\begin{align*}
\theta^{\Hh}_{\Qo} \Do &= (q_n^{\theta}) (r_n) Sw \So \\
\theta^{\Hh}_{\Qo} \So &= (q_n^{\theta}) (r_n^{-1}) Sw \Do
\end{align*}
where $q_n^{\theta}$ is the quaternionic value of $\theta^{\Hh}_{\Qo}$, and $r_n = (q_{n d})(q_{n s})^{-1}$. 

The result is that we get new Dirac-Weyl-Higgs equations for the fermion fields:
\textbf{Quarks and Leptons}:
\begin{align*}      
(\DbhL - \half iI\dtcl - \half iI\dthl) q_L &= (\Higgs  r_n Sw) q_R \\
(\DbR - \half iI\dtcr - \half iI\dthr) q_R &= (\Higgs r_n^{-1} Sw) q_L \\
(\DhL - \half iI\dtcl - \half iI\dthl) l_L &= (\Higgs r_n Sw) l_R \\
(\DR - \half iI\dtcr - \half iI\dthr) l_R &= (\Higgs r_n^{-1} Sw) l_L
\end{align*}

\textbf{Anti-quarks and Anti-leptons}:
\begin{align*}      
(\DbLc + \half iI\dtcl + \half iI\dthl) \bar{q}_L &= (\Higgs r_n^{-1} Sw) \bar{q}_R \\
(\DbhRc + \half iI\dtcr + \half iI\dthr) \bar{q}_R &= (\Higgs r_n Sw) \bar{q}_L \\
(\DLc + \half iI\dtcl + \half iI\dthl) \bar{l}_L &= (\Higgs r_n^{-1} Sw) \bar{l}_R \\
(\DhRc + \half iI\dtcr + \half iI\dthr) \bar{l}_R &= (\Higgs r_n Sw) \bar{l}_L
\end{align*}

We now have pairs of coupled equations with all terms confined to the tensor product.  We shall now show that this is isomorphic to the Standard Model Dirac-Higgs equations for quarks and leptons, with the physical hypercharge connection, Higgs field and mass matrices. The first step is to derive the Standard Model hypercharge from our tensor product of theta connections.
      
\subsection{Derivation of Hypercharge}\label{S:tenshyp}

So far, we have derived Dirac-Higgs equations from our harmonic map tension equation that are very similar to their Standard Model equivalents, but we have only analyzed the irreducible representations of the $SU(3)Sp(1)$ part of the gauge connections. The Standard Model gauge group is $SU(3)SU(2)U(1)$, where $U(1)$ defines what the physicists call ÒhyperchargeÓ. In this section we will show how our ÒhypercolorÓ operator and ÒhyperspinÓ operator tensor together to produce the ÒhyperchargeÓ $U(1)$ gauge interaction.
 
\subsubsection{Hypercolor}

We already know quite a bit about the $\D \otimes \half iI\dtc$ hypercolor operator. Its $U(1)U(1) \cong U(1)$ holonomy action is on $V \cong \mathcal{O}(-2)_{\CP^3}$ via the $V_{-2}$ representation of $U(1)$. Let us therefore define the ÒhypercolorÓ $Y_C$ of the line bundle $\mathcal{O}(n/2)$ representations of $\D \otimes \half iI\dtc$ as $n/2$. So the hypercolor of the $V$ component of any tensor is $Y_C(V) = -2$. Switching to $\bar{V} \cong \mathcal{O}(+2)$, leads to a hypercolor of $Y_C(\bar{V}) = +2$.
      
The operator $\D \otimes \half iI\dtc$ also acts on $\Th\CP^3_o \cong T$ via a representation of $U(1)$. If we say that it acts the same on all subspaces of the tangent space, the lie algebra representation has the form of a pure imaginary number multiplied by the $3 x 3$ identity matrix, say 
\[
\begin{pmatrix} 
a & 0 & 0 \\ 
0 & a & 0\\
0 & 0 & a 
\end{pmatrix}
\] 
The trace of this number is the hypercolor of the operator on the particular line bundle associated with the tangent bundle by the mapping of $T \to \Th\CP^3_{Io}$ and $V \to \mathcal{O}(-2)$. Since, as we shall learn in the next section on hyperspin, this same map defines a complex contact bundle $\mathcal{O}(+2)$, on $\CP^3_{Io}$, that associated bundle must be $\mathcal{O}(+2)$, whose hypercolor is obviously $+2$.
So, clearly $3a = +2, \quad a = +2/3$, and thus
\[
Y_C(T) = +2/3
\]
Clearly also $Y_C(\bar{T}) = -2/3$.

\subsubsection{Hyperspin}\label{S:tenshypspin}

The hyperspin operator $\dd \otimes \half i_nI_n \dth$ is quaternion-valued. In order to understand its releationship to hypercharge, we need to reduce its quaternionic action on the quaternionic line bundle $S_n$ to that of a complex connection two different complex line bundles. In order to do this, we need to single out a complex structure, $I$. This will allow us to fiber over $\CP^3_{Io}$, as we did in the LCK case, instead of over $\HP^1_o$.  This will have the effect of reducing the $Sp(1)Sp(1)$ hyperspin operators to complex-valued $U(1)U(1)$ operators similar to the hypercolor: $\D \otimes \half iI_n\dtc$, with only the different complex structures $I_n$ to differentiate it from hypercolor.

This specification of a particular $I$ is accomplished by the tensor product. Tensoring $A_n \xc \Th$ and $\bar{A}_n \xc \Tah$ singles out a particular $i \in S^2$ in the one-dimensional quaternionic subspaces at each point defined by $S_x, D_x$, making each subspace isomorphic to $\C^2$. Thus it defines one complex structure $I \in S^2$ in the hypercomplex structure on \ML. The $Sp(1)_{iso}$ holonomy group can be identified with $SU(2)_I$. Under this isomorphism 
\[
A_n\ML \cong \Th\ML  \quad \bar{A}_n\ML \cong \Tah\ML
\]

Using $I$, we can define several additional distributions on $A_n$ and $\bar{A}_n$, besides the standard distributions $S_n, \bar{S}_n, D_n, \bar{D}_n$. We use the mappings $(1 - iI), (1 + iI)$ to map the distributions $T = ker(\theta + I\theta)$ and $V = \langle \theta^\#, I\theta^\# \rangle$ into holomorphic and antiholomorphic distributions:
\[      
 T_{n I}, \bar{T}_{nI} := (1 \mp i I_n)T \quad V_{nI}, \bar{V}_{nI} := (1 \mp i I_n)V
 \]      

Since $T_{nI}$ holomorphically maps onto $\Th\CP^3_o$, we can use $D_n$ to define a complex contact distribution on $\CP^3_{Io}$. The contact line bundle of this complex contact structure is then isomorphic to a one-complex dimensional subdistribution of $S_{nI}$ that we will call $L_{nI}$. According to Salamon, for any complex structure, this contact line bundle on $\CPI^3$ is $\mathcal{O}(+2)$ \cite{ss82}. Thus, the complex structure, $I$, is determined by the tensor product with LCK bundles, but the complex \emph{contact} structure on $T_{nI} \cong \Th\CP^3_o$ is determined by the rest of the hypercomplex structure, which defines $D_{nI}, S_{nI}$. Clearly, 
\[
T_{nI} = D_{nI} + L_{nI}
\]
Also, note that 
\[
S_{nI} = V_{nI} + L_{nI}
\] 

Thus we can redo our LCHK reduction, this time going to the complex distributions $T_{nI}$ and $V_{nI}$, instead of the quaternionic $S_n$ and $D_n$. This means that the effect on the pullback equations of reducing the LCHK equation to $T$, that is, to $\Th\CP^3_o$, is to split the quaternionic bundle $S_nI$ into two complex line bundles $V_{nI}$ and $L_{nI}$. This is equivalent to reducing the $Sp(1)Sp(1)$ gauge group of $\dd \otimes \half i_nI_n \dth$ into a $U(1)U(1)$ gauge group of $\D \otimes \half iI_n\dtc$. (Notice the analogy with reducing the $U(1)SU(2)$ electro-weak interaction group to a $U(1)U(1)$ group generated by the hypercharge and the third isospin component, $T_3$.) 

Since the Weyl connection preserves the hypercomplex structure, it also preserves this complex contact structure. The key difference, when reduced to the tangent bundle of $\CP^3_{Io}$, between the LCK Weyl connection and the LCHK Weyl connection, is that the LCHK holonomy preserves the complex contact structure, whereas the LCK holonomy does not. The holonomy group differences between the LCK and LCHK cases can be seen to be the result of the different homogeneous space realizations of $\CP^3$ as either $SU(4)/U(3)$ or $Sp(2)/U(2)$, with well-understood implications for homogeneous kaehler metrics. The holonomy of the hyperspin connection is intimately bound up with the complex contact structure.

The holonomy representations of the connections on $T_{nI} = D_{nI} + L_{nI}$ can be understood by considering the decomposition of the isotropy representation of $Sp(1)U(1)$ on the homogeneous complex contact space $\CP^3_I = Sp(2)/SU(2)U(1)$, with an $\mathcal{O}(+2)$ contact line bundle, namely:
\[      
      D_{nI} + V_{nI} = V_1(Sp(1)) \otimes V_0(U(1)) + V_0(Sp(1)) \otimes V_2(U(1))
\]      
That is,  the $SU(2)$ connection acts via the $V_1$ representation on $D_I^{1,0}$, while the $U(1)$ connection is trivial ($V_0$). The $SU(2)$ connection is trivial on $V_I^{1,0}$, while the $U(1)$ connection acts via $V_2$. 

We will define the $U(1) \cong U(1)U(1)$ representation of $\D \otimes \half iI_n\dtc$ as the Òhyperspin,Ó $Y_S$. Since$V_{nI} \cong \mathcal{O}(-2)$, for all $n$, and $L_{nI} \cong \mathcal{O}(+2)$, for all $n$, the effect of $D \otimes \half iI_n\dtc$ on these line bundles is the same as $D \otimes \half iI\dtc$:
\[
Y_S(V_{nI}) = -2, \  Y_S(L_{nI}) = +2, \  Y_S(\bar{V}_{nI}) = +2 ;\  Y_S(\bar{L}_{nI}) = -2.
\]    
       
Since $D \otimes \half iI_n\dtc$ is trivial on $D_{nI}$, that means that $Y_S(D_{nI}) = 0$, and also $Y_S(\bar{D}_{nI}) = 0$. The term ÒhyperspinÓ is chosen as a deliberate nod to the Standard Model term ``third component of weak isospin,'' $T_3$, Hyperspin is not defined the same way as $T_3$, but it has exactly the opposite effect, since $iI_n\theta_{\Hh}$ is the third component, not of weak isospin, which is the isotropy group $Sp(1)_{iso}$, but of the other $Sp(1)$ subgroup of $Sp(2)$. Hyperspin is trivial on $Sp(1)_{iso}$ doublets, but non-trivial on $Sp(1)_{iso}$ trivial singlets.

\subsubsection{Hypercharge and a Comparison with Physics}

The tensor product connection $\nabla_{U(1)} \otimes iI\theta \otimes iI_n\theta$, combining the hypercolor connections and the hyperspin connection, can be computed straightforwardly. In this case, the hyperspin and hypercolor quantum numbers can add together to produce a single number.

The distributions $T$ and $D_{nI}$ are both in $\Th \ML$, and $V$ and $V_{nI}$ have both been identified as isomorphic to $\mathcal{O}(-2)$ over $\CPI^3$. Similarly, we established that $L_{nI} \cong \mathcal{O}(+2)$. The tensor products of line bundles are easy to compute, e.g
\begin{align*}      
     V \xc V_{nI} &\cong  \mathcal{O}(-2)\mathcal{O}(-2) = \mathcal{O}(-4) \\
     V \xc L_{nI} &\cong  \mathcal{O}(-2)\mathcal{O}(+2) = \mathcal{O}(0)
\end{align*}	      

So if we define the connection $\nabla_{U(1)} \otimes iI\theta_{\C} \otimes iI_n\theta_{\Hh}$ on these product bundles, the product representations will be parameterized by $2Y := Y_C + Y_S$, where $Y$ is the Standard Model ÒhyperchargeÓ. Hypercharge is thus the tensor product of the two theta connections. 

Thus by employing the standard definition of electromagnetic charge $Q = T-3 + 1/2Y$ we get Table \ref{T:charges}, containing the fermion hypercharges and charges derived by our model.  It is identical to the corresponding hypercharges and charges of the Standard Model for each generation.
\begin{table}\label{T:charges}

\begin{align*} 
&Fermion	&      &Tensor & Y_C&    &        Y_S&     &        2Y&    &   4T_3&       &  Q& \\
\hline
\\
&(d/u)_L 	&	   &TD_{nI}    & +2/3&       &      0&   &	  +2/3& 	&          (-2, +2)&   &  -1/3, +2/3& \\
&(e/\nu)_L 	&      &VD_{nI}    & -2&		    &      0&   &		-2&	    	&		 (-2, +2)&   &     -1, 0& \\
\hline 
\\
&d_R, u_R 	&	&TV_{nI}, TL_{nI}  & 	+2/3&  &  (-2, +2)&  &   -4/3, +8/3&  &   0&	 &   -1/3, +2/3& \\
&e_R, \nu_R  &   &VV_{nI}, VL_{nI} &	-2&	  & (-2, +2)&	&	-4, 0&	      & 	0&		&	-1, 0& \\
\hline
\\
&\bar{d}_L, \bar{u}_L &	&\bar{T}\bar{V}_{nI}, \bar{T}\bar{L}_{nI} &	-2/3& &	(+2, -2)&	& 	-4/3, +8/3&  &   0&	  &  +1/3, -2/3&\\
&\bar{e}_L, \bar{\nu}_L &  &\bar{V}\bar{V}_{nI}, \bar{V}\bar{L}_{nI} &  +2&  & (+2, -2)&	&	-4, 0&  	   &   0&	&	+1, 0& \\
\hline
\\
&(\bar{d}/\bar{u})_R 	&	&\bar{T}\bar{D}_{nI} &	-2/3&  &	0&	&	-2/3&	&     (+2, -2)&  &   +1/3, -2/3& \\
&(\bar{e}/\bar{\nu})_R &	&\bar{V}\bar{D}_{nI} &	+2&   &	0&	&	-2&	    	&	(+2, -2)&	&	+1, 0& \\
\hline
\end{align*} 
\caption{Hypercharge $2Y$ = Hypercolor $Y_C$   + Hyperspin $Y_S$}
\end{table}

\section{Statement of the Hypothesis and Additional Support}\label{S:hyp}

At this point, let us pause and take stock. We have been examining the geometry of \ML, the moduli space of hyperholomorphic \Hh - line bundles on the hyper-elliptic Hopf surface.  Such Hopf surfaces are like ``hyperelliptic quaternionic curves'' and are in many ways the quaternionic analogues of elliptic curves in complex geometry. The moduli space itself is the quaternionic analogue of the Picard variety of an elliptic curve, a very important structure in the study of the properties of these curves.  Thus we are studying a moduli space that should be very important for the study of four-dimensional manifolds in general, since quaternions seem to play a role in the very unusual properties of four dimensions.
      
Now, one four-dimensional manifold is the space-time of classical field theory. So one might expect $\ML$ to have some relevance to physics. But the results we have obtained are quite extraordinary. The two complementary aspects of the Hopf surface, its quaternionicity and its ellipticity led to quaternionic constructions, such as the LCHK structure and Swann fibration, and also to complex constructions, such as the LCK structure and $SL(2,C)$ spin connection. These have led us to unique holomorphic sections of the fiber bundle $\ML \to \HP^1_o$, each defined by one of a very restricted class of Lorentzian metrics on $\HP^1_o$. Because of the natural structures we have defined, these sections are harmonic maps from $\HP^1_o \to \ML$, and thus solve Euler-Lagrange equations. These Euler-Lagrange equations (the tension equation), turn out to be identical to the unbroken Standard-Model Dirac-Higgs equations for all known fermion fields, all known interactions, the precise quantum numbers for the interactions.
      
      One is led to suspect that there is something profoundly important to physics about \ML.  We will therefore propose the following very strong hypothesis:
            
\begin{hypothesis}
Physics is the geometry of \ML: classical field theory is its differential geometry; quantum field theory is its algebraic geometry.
\end{hypothesis}

In this section \ref{S:hyp}, we shall provide some additional support for this hypothesis. 

First, in section \ref{S:curve}, we will examine the other half of the equations that comprise the Standard Model, namely, the Yang-Mills equations solved by the curvatures of the interaction connections, which couple to ÒcurrentsÓ constructed from the fermionic fields.  We will find that, here too, $\ML$ provides exact counterparts to the Standard Model. We will also show that the equation solved by the Higgs field is itself a form of Yang-Mills equation for a curvature, and this will illuminate the role of the Higgs mass parameter as another interaction coupling constant. We will find that all four coupling constants (including the Higgs mass) are derived from an unbroken Òcoupling fieldÓ consisting of the Levi-Civita connection form of the Lorentzian metric.

The remaining subsections are quite brief. Each deals with a very large topic and is meant to provide only a sketch of an argument that supports our hypothesis.
      
In section \ref{S:grav}, we will examine some of the implications of this construction for the Lorentzian metric itself.  We will find that since all constant-parameter quadratic terms have been removed, all fields are massless and the metric is conformally flat.  Thus, the unbroken equations (which presumable describe a very high-temperature regime indeed!), describe a metric of the sort proposed by Penrose for the very early universe.
      
In section \ref{S:higgs2}, we will revisit the Higgs field, fermion generations, and the mass matrices. We will find that the 20 components of the mass matrices, CKM matrix, and PMNS matrix, plus the four components of the Higgs field itself, are consistent with defining the orientation of $E$ relative to $D_n$ and $S_n$, and thus of the original choice of Lorentzian metric $g$, and thus spinor $\Ps$, complex structure $I$, and section $\PL$. 
      
In section \ref{S:qft}, we will look at quantization. As is typical in non-linear sigma models, we treat the sections $\PL: \HP^1_o \to \ML$ as the coordinate functions of \ML. We then construct a state space by taking symmetric (in the LCHK case, antisymmetric) local products of these coordinates. Since our maps are holomorphic, this state space is equivalent to the sheaf of holomorphic functions on $\ML$ in the LCK case, and a somewhat more elaborate sheaf in the LCHK case. The harmonic maps act as coordinate fields, while the $\Zo$ and $\Qo$ fields act as derivation operators on this sheaf. When we tensor together the fields, the resulting Òfield operatorsÓ obey the standard commutation and anti-commutation rules of quantum field theory. In addition, the original Dirac field on $\HP^1_o$, defined by the parallel spinors of the Lorentz metric, lifts to define a holomorphic flow on \ML. The QFT version of that flow defines sheaf homomorphisms of state space sheaf that are consistent with the QFT time evolution operator.

\subsection{Curvature Equations and Currents}\label{S:curve}

So far, we have taken the Weyl connections on $\ML$ as fixed and asked how their configuration governed the harmonic sections $\PL$. In the terminology of physics, we have assumed an ÒexternalÓ force field.  The Standard Model, however, has the interaction connection and the fermion fields coupled together.  Just as the fermion fields solve the connection-defined Dirac-Higgs equations, so the connections (their curvatures, actually) solve the Yang-Mills equations which couple them to ``currents'' constructed from the fermion fields. Physicists say that the fermion fields generate the gauge (curvature) fields, just as the gauge (connection) fields reciprocally control the fermion field's behavior.
      
We can interpret the tension equation as describing how the distribution $E$ looks in terms of structures on \ML.  The Yang-Mills approach is the opposite: to describe how the structures on $\ML$ look in terms of the distribution $E$. The Yang-Mills equations have the general form 
\[      
      tr_g (\nabla \circ F^\nabla) = c_\nabla j(\psi_f)
\]      
where $tr_g$ is the same trace operator that we encountered in the tension equation, $F^\nabla$ is the curvature of the connection $\nabla$, $j(\psi_f)$ is the lie-algebra-valued current associated with the fermion field $\psi_f$, and $c_\nabla$ is the ``coupling constant'' associated with the ``interaction'' $\nabla$. Since $\wk{}{}, \whk{}{}$ are defined on \ML, while $g$ exists on $\HP^1_o$, we have to use the lifted ÒcometricÓ $g*$ on the distribution $E$ in order to perform the trace, as before.  Since the lifted (sub)metric $g$ is (sub)kaehler on $E$, we know that
\[
	tr_g (\nabla \circ F^\nabla) = \nabla (F'(g^*))
\]
where $F'^\nabla$ is the pullback to $E$ of the curvature tensor $F$, and we take a Ricci-style trace, except that the result is a Lie Algebra-valued function, instead of a scalar-valued two-tensor. (Kobayashi discusses $F'(g^*)$, which he calls the mean curvature tensor, and the Ricci tensor in depth in \cite{sk87}, and shows their equivalence on kaehler manifolds. Besse \cite{ab87} discusses exchanging $tr_g$ and $\nabla$ for kaehler metrics.) 

The connections we defined in previous sections were pulled back to $E$ from connections on $\ML$ that were themselves lifted (after our reduction process) from various standard vector bundle connections on the kaehler manifold $\CP^3_{Io}$. All these connections are themselves pulled back from connections that are homogenous under the isometry groups of the kaehler symmetric space $\CPI^3$. 

The bundles involved are isomorphic to the tangent bundles $\Th \CPI^3, \Tah \CPI^3$ and the line bundles $\mathcal{O}(-2), \mathcal{O}(+2)$. In what follows, we shall use the symbol, $\nabla$,  for any one of the many connections on these bundles that we defined in previous sections and pulled back to $E$ and to $\HP^1_o$.

We know from Opferman and Papadopoulos \cite{ao08}, that in the homogeneous case the curvature form and symmetric curvature tensors of these homogeneous connections take on a very simple form:
\begin{align*}      
      F^\nabla &= \half Y^i ([Y^i]_{mn} dx^m \wedge  dx^n)\\      
      F'^\nabla &= \half Y^i ([Y^i]_{mn} dx^m \otimes dx^n)
\end{align*}      
where $Y^i$ are the generators of the lie algebra of the isotropy group of the homogeneous connection (which is the holonomy group), and $dx^m$ is a co-frame of forms on$\CPI^3$. After combining our expressions from above, and writing the $g$-trace on $E$ explicitly, we get:
\begin{align*}
	tr_g(\nabla F'^\nabla) &= \nabla ( \half Y^i ([Y^i]_{mn} dx^m(X^\mu) \otimes dx^n(X^\mu)))\\
			&= Y^i([Y^i]^{mn} dx^m(\lcl{}{X^\mu}) \otimes dx^n(X^\mu))
\end{align*}
where $\lcl{}{}$ is the Levi-Civita connection for $g$, and we know that on symmetric spaces $\nabla(F^\nabla) = 0$.

We will be looking at tensor products of vector fields with parallel spinors, of course, so we will have to evaluate that expression carefully. The expression on the right is a lie-algebra-valued symmetric 2 form, so the trace of the comparable form for the curvature of our connections on \ML, e.g., $\wk{}{}$, tensored with the Lorentz spin connection, $\spl{}{}$, includes a hermitian inner product, $\langle, \rangle$, on spinors:
\[      
	= Y^i([Y^i]_{mn} dx^m(\lcl{}{X^\mu}), dx^n(X^\mu) \langle \PsL, \PsL \rangle
\]      
where, as always, $\PsL$ is a parallel spinor, left in this case. Continuing:
\[
	 = Y^i([Y^i]_{mn} dx^m(\lcl{}{X^\mu}), X^0) \langle \PsL, \sig  \PsL \rangle
\]
where we have use the properties of Jordan multiplication to move $X^\mu$ to the spinor side. Moreover, by the definition of the Levi-Civita connection acting on the frame fields $X^\mu$, we have:
\[
	 = a_X Y^i dx^\mu ([Y^i]_{mn} dx^m(X^0), dx^n(X^0)) \langle \PsL, X^\mu . \PsL \rangle
\]
where the real scalar field $a_X$ is defined as
\[
	a_X := \sum_{vk} X^\nu \Gamma_{\nu\kappa}^0 X^\kappa
\]
where $\Gamma_{\nu\kappa}^\mu$ are the Christoffel symbols for $\lcl{}{}$.

We will call $a_X$, the unbroken Òcoupling field.Ó It corresponds to the coupling ÒconstantsÓ $g, g_2, g_1$ of the Standard Model, which at high energy would then converge to a single value.  In this theory, the coupling field is derived from the gravitational connection on the distribution $E$. If the pp-wave gravitational metric, $g$, has a non-zero minimum-energy solution, as the Higgs field has, then at ``low'' temperatures, the coupling field could freeze out as a coupling constant. This observation would seem to reinforce the notion that our Dirac-Higgs equations describe a high-temperature regime.

Writing our currents in a more compact form, here are a few representative samples:
\begin{align*}      
      a_X \langle q_R, [Y^i] \sig \cdot q_R \rangle 		&&\text{right-quark color current}  \\   
      a_X \langle l_L, [Y^j] \sig \cdot l_L \rangle			&&\text{left-lepton weak current}      
\end{align*}      
where $[Y^i], [Y^j]$ are the generators of the $SU(3), Sp(1)$ groups, and we have only shown the coordinate representation of the currents. The hypercharge currents are the sum of  hypercolor and hyperspin currents, each of whose lie algebra is generated by $i$.

Clearly the quaternionic Higgs field as we have defined it, $\theta^{\Qo}$, is part of the LCHK Weyl connection, and thus has a curvature
\[      
     d_{X^\mu} - \theta_{\Qo} \theta_{\Qo}
\]      
where the action of $\theta_{\Qo}$ on itself is by quaternionic multiplication. Some of the currents associated with this curvature are
\begin{align*}
	a_X \langle l_R, [Y^j] \sig \cdot l_L \rangle 		&&\text{leptonic Higgs current}\\
	a_X \langle \bar{q}_R, [Y^j] \sig \cdot \bar{q}_L \rangle 	&&\text{anti-quark Higgs current}
\end{align*}
The dimensionless ÒmassÓ coupling constant of the Higgs interaction $\half (m_h / v)^2$ is clearly the parameter that corresponds to $g, g_1, g_2$ in the other curvature equations. In our theory, it too is our coupling field $a_X$.

 This examination of curvature and currents has shown that the $\ML$ equations are essentially identical to the Yang-Mills equations for gauge fields (curvatures) with interacting with currents and a coupling field. We also suggest that the Higgs coupling to currents is also such a Yang-Mills equation. Our strong hypothesis seems to be well supported.

\subsection{Einstein Equations}\label{S:grav}

We defined our Lortentzian metric $g$ as supporting parallel spinors. This implies that it is a Òpp-waveÓ or ÒBrinkmannÓ metric. In addition, by requiring that the complex structure on the twistor space be integrable, we are implying that $g$ is conformally flat.  Conformally flat, pp-wave metrics are known to provide exact solutions to the Einstein gravitational equations, but they are so-called Ònull dustÓ solutions, meaning that they describe a universe with only massless fields and gravity.
      
At first this may seem like an unreasonable assumption to make about the universe, but in the case of our equations, it makes sense.  Our coupled equations are completely massless.  All the mass and coupling Lagrangian parameters (the ones that break conformal symmetry) have been replaced by scalar fields. So the equations do describe a conformally flat universe filled with massless fields and gravity.
      
Penrose suggested that such conformally flat metric might describe the early universe just after the Big Bang \cite{rp79}.  We suggest that our unbroken equations, with non-minimum coupling and Higgs fields, describes a very high-temperature conditions, and that at lower temperatures, the Higgs and coupling fields freeze out to their minimal energy values, and the observed universe takes on its familiar form. One could also say that these equations describe a very high-energy situation in general. When we discuss quantization, in section \ref{S:qft}, we will revisit some of these issue.
      
Our harmonic maps solve a minimization problem that includes both $g$ and the $\ML$ metrics, so that minimization problem must be equivalent to one including our version of the fully unbroken classical Standard Model along with the pp-wave gravitational metric. Thus the Eells-Lemaire stress-energy two-tensor of the harmonic map \cite{je83} is
\[      
      S_\psi = e_\psi g Ð \psi_{g\lambda}^*b
\]      
where $e_\psi$ is the energy density of the map, and $\psi_{g\lambda}^*$ is the pullback of the LCK or LCHK metric. Since we know that the Einstein equation works for pp-wave metrics, the traceless Ricci tensor (or Einstein tensor) must be proportional to $S_\psi$, with the proportionality constant $G$, namely the gravitational constant. Since the Einstein tensor includes $g$, this suggests that
\[
	e_\psi^{-1}S_\psi = g - e_\psi^{-1}\psi_{g\lambda}^*b = g - \psi_{g\lambda}^*b/tr_g(\psi_{g\lambda}^*b)
\]
is $G S_\psi$ in the Einstein equations. (We have used the shorthand $tr_g(\psi_{g\lambda}^*b)$ for $||d\PL||^2_{bg}$, because it shows more clearly the relationship of the energy density to $g$ and $b$. And we have neglected $h$ in this discussion, but it is there also in the complete expression.)	
	
Thus, in this model
\[
	G = e_\psi^{-1}
\]
that is, the gravitational ÒconstantÓ is just the inverse of the local energy density the harmonic map that defines our particular universe. Obviously, this value may vary very slowly, so it may appear constant, but in this theory can vary.

The hypothesis still holds up pretty well, although in its classical unbroken form it seems to describe only the hot, very early universe or very high-energy interactions.

\subsection{Higgs Fields, Generations, and Mass Matrices}\label{S:higgs2}

When we left the Higgs, field in section \ref{S:higgs}, it had the form
\[
\Higgs r_n Sw
\]
where at any point
\begin{align*}
\theta^{\C}_{\Zo}(x) &:= \theta(\Zo)(x) = \theta(X^0 - iIX^0)(x) \in \C \\
\theta^{\Hh}_{\Qo}(x) &:= \theta(\Qo)(x) = \theta(X^0 - i_n I_n X^0)(x) \in \Hh^3 \\
r_n(x) &:= (q_{n d})(x)(q_{n s})^{-1}(x) \in \Hh^3
\end{align*}
where $q_{n d}, q_{n s}$ are the quaternionic magnitudes of the fields $\Do, \So$. Neglecting $Sw$, whose form is fixed by the frame, as described in \ref{S:higgs}, and for the moment also the complex Higgs field $\theta(\Zo)$, the quaternionic Higgs field and the $r_n$ field each have twelve real parameters (three quaternionic parameters each). 

It is plausible that the twelve $r_n$ parameters are related to the twelve real mass values in the diagonal mass matrices of the Standard Model. Each generation has four mass matrix values: $m_\nu, m_e, m_u, m_d$, which would then represent a single quaternion. In the Standard Model these values are fixed, in this model, they constitute a field that specifies at any point the quaternionic orientation of the $E$ distribution in $A\ML$. The Standard Model 3 x3 real mass matrices for each particle type would then be combined into a single 3 x3 quaternionic matrix.

In the Standard model, the other twelve real values associated with the Higgs field, are (a) two complex numbers specifying the Higgs doublet, (b) four real values for the CKM matrix, and (c) four real values for the PMNS matrix. We hypothesize that these three values are each quaternionic, and are represented in this model by $\theta(\Qo)$, the quaternionic Higgs field. In other words, in this model, the Higgs field is tri-quaternionic and includes the CKM and PMNS values. 

The complex Higgs field, $\theta(\Zo)$, measures the complex magnitude of the $\Vo$ field, in other words, the leptonic component. Presumably, it enters the picture by inducing the split between the CKM (quark) matrix and the PMNS (lepton) matrix.

In the Standard Model, the Higgs field has all four degrees of freedom at high temperatures, but freezes to single fixed value at a transition temperature. On the other hand, in the Standard Model, the mass, CKM, and PMNS matrices are fixed. Presumably, if this model is correct, then either the parameters are very slowly changing and reflect the local orientation of $E$, or else they have frozen out at some even higher temperature.  There is thus a parallel between our treatment of the quaternionic Higgs field, the coupling field, and the gravitational constant.

\subsection{Algebraic Geometry of $\ML$ and Quantum Field Theory}\label{S:qft}

Since $\ML$, is a manifold, it is a pair, $(\ML, \mathcal{S})$, where $\ML$ is the underlying topological space and $\mathcal{S}$ is the structure sheaf of functions over open sets of $\ML$. Basically, the structure sheaf is generated by the coordinate functions of the manifold, so for complex manifolds, say, the holomorphic coordinate functions generate the structure sheaf, $\mathcal{O}_{\ML}$, of holomorphic functions on $\ML$. The great insight of algebraic geometry is that the \emph{geometric} properties of the manifold $\ML$ can be derived from the \emph{algebraic} properties of its structure sheaf. We will now begin to apply this insight to our model, finding very quickly that the result is to ``quantize'' our classical field theory.

The idea that the ÒquantumÓ version of a ÒclassicalÓ geometry is based on the set of functions on the manifold, instead of maps into the manifold, is by now fairly common. Berger's monumental tome from 2003 \cite{mb03}, for example, analyzes Riemannian manifolds from two viewpoints: as Òquantum mechanical worlds,Ó via spectral geometry, and as Òdynamical systems,Ó meaning classical worlds, via geodesics. Witten, in his influential paper on Morse theory \cite{ew82}, pointed to the close similarity between the ideas of quantum field theory and the properties of Morse functions on manifolds. And of course the theory of geometric quantization uses the sections of a line bundle over the symplectic manifold of classical states to quantize Hamiltonian mechanics. We will be focussing specifically on the structure sheaves of functions of $\ML$ that define its complex structure and hypercomplex structure.

We will find that the germs of functions in these sheaves represent quantum states, and that our classical fields naturally act as operators on these germs in the precise fashion that Standard Model field operators act on the states of the associated quantum field theory.  We will also find that the most important operator on quantum states, the interaction propagator (also called the ``time-evolution operator of the interacting field theory'' \cite{mp95}), is induced by a natural holomorphic flow on $\ML$ defined by our classical equations. By contrast to some previous theories, our model does not require the construction of state spaces over infinite dimensional manifolds of classical solutions. Instead, we will only use sheaves over the finite-dimensional manifold \ML.

\subsubsection{Complex Structure Sheaf on  \ML} \label{S:comsheaf}

Since $\ML$ is a complex manifold, it is also a complex analytic space, with structure sheaf of holomorphic functions, $\mathcal{O}_{\ML}$. These functions are generated by the holomorphic coordinate functions $z^i: \ML \to \C$. At each point $p$ of $\ML$ the \emph{stalk} $\mathcal{O}_p$ of the sheaf is comprised of the \emph{germs} of holomorphic functions at $p$. 

Since $\mathcal{O}_{\ML}$ is a commutative ring, the individual stalks are also commutative rings, thus germs can be multiplied by each other and added together. Since the coordinate functions themselves are elements of the sheaf (linear functions), germs of the coordinate functions act as operators on stalks by germ multiplication, which is symmetric in this case.  In fact, the entire stalk, $\mathcal{O}_p$, is generated by the germs of the coordinate functions at $p$, through symmetric multiplication and addition. 

As germs of linear functions they are representatives of cosets in $\mathfrak{m}_p / \mathfrak{m}^2$, where $\mathfrak{m}_p$ is a maximal ideal of $\mathcal{O}_{\ML}$, consisting of the holomorphic functions that vanish at $p$. The space $\mathfrak{m}_p / \mathfrak{m}^2$ is defined as the cotangent space to $\ML$ at $p$. There are thus natural mappings from germs of coordinate functions into both covectors and operators on $\mathcal{O}_p$.

Similarly, tangent vectors are defined as derivation operators on $\mathcal{O}_p$, that is sheaf homomorphisms that follow the Leibnitz rule.Thus derivation operators on germs $\phi_p, \beta_p$ at $p$ are defined by 
\[
D_p(\phi_p \cdot \beta_p) = (D_p\phi_p) \cdot \beta_p + \phi_p \cdot (D_p\beta_p)
\]
where we use $\cdot$ to represent ring multiplication in the stalk. Thus, both the germs of holomorphic coordinate functions $z^i_p: \ML \to \C$ and the holomorphic tangent basis vectors at a point, $Z^i_p$ act as operators on the stalk $\mathcal{O}_p$.

\subsubsection{Hypercomplex Structure Sheaf on \ML}\label{S:hypsheaf}

$\ML$ is also a hypercomplex manifold, but there has been very little written on the appropriate structure sheaf to use for a hypercomplex manifold. Verbitsky \cite{mv97}, following a suggestion by Simpson and Deligne \cite{cs91}, defined a hypercomplex variety, $M$, as a variety having a sheaf of complex differentials, $\Omega^1(\mathcal{O}_M)$ with three separate complex structures and thus a natural $SU(2)$ action on the sheaf. 

We will take a leaf from Widdows' \cite{dw00}, and define the structure sheaf of a hypercomplex manifold by reference to the three bundles of complex differentials we used earlier, \ref{S:tenshypspin}. 
\[
\Omega^1_n : = \Gamma (\ML, \Lambda_n^{1,0}) := \Gamma (\ML, (1+ i I_n)T^*) 
\]
Widdows noted that the three $\Lambda_n^{1,0}$ spaces represent a single $sp(1) = su(2)$-valued $\Lambda^{1,0}$ space, with $Sp(1) = SU(2)$ acting via the adjoint representation on $sp(1)$. So with that as a suggestion, we will define the coordinate functions of the hypercomplex manifolds as the following functions on the atlas $\{U_{\alpha}\}$
\[
q^i_n : U_{\alpha} \subset \ML  \to  \C \otimes sp(1)
\]
and construct the full sheaf by successive ring multiplications and additions of the coordinate functions, as we did with $\mathcal{O}_{\ML}$. Multiplication will involve ordinary complex multiplication in the $\C$ component, combined with the lie algebra bracket operation in the $sp(1)$ component. The new element is that this multiplication is anti-commutative:
\[
q^i_n \cdot q^j_m = - q^j_m \cdot q^i_n
\]

We will call the resulting sheaf $\mathcal{A}_{\ML}$. As with $\mathcal{O}_{\ML}$, both germs of coordinate functions, $q^i_{np}$ and tangent vectors, $Q^i_{np}$ act as operators on the stalks $\mathcal{A}_p$.

\subsubsection{Classical and Quantum States, Fields and Field Operators}

We will define a ``classical state,'' as a pair 
\[
(p, (Z \otimes_{\C} Q_n)_p) \in (\ML, \Th\ML \otimes_{\C} A_n\ML)
\]
For every classical state there is a stalk $(\mathcal{O} \otimes \mathcal{A})_p$ of germs, and all the germs in the stalk are ``quantum states'' that are associated with that classical state.  Note that quantum states at different points can be related to each other by taking global section of the sheaf. However, as is well known, the problem of finding global sections of a sheaf over a manifold is usually non-trivial and always intimately related to the geometry of the manifold. 

Let us define the germs of the constant sheaf 
\[
\C_{\ML} \otimes_{\C} \C_{\ML} \subset \mathcal{O}_{\ML} \otimes_{\C} \mathcal{A}_{\ML}
\]
as the ``vacuum state'' on each stalk, $(\mathcal{O} \otimes_{\C} \mathcal{A})_p$. 

``Multi-particle states'', are then constructed from the vacuum by the ``creation operators'' $z^i \otimes q^j_n$. For example the $z^i \otimes q^j_n\C$ are ``one particle states.'' This process can continue until the entire $(\mathcal{O} \otimes \mathcal{A})_p$ stalk is generated by repeated multiplications and additions and scalar multiplications. Similarly, operations by vectors, the ``annihilation operators'', subtract one particle from the state. For example $Z^i (z^i \C) = \C, \quad Z^i \C = 0$. Depending on which pair of creation operators is chosen, we can obviously create all our particle types: quark and anti-quark doublets and singlets, lepton and anti-lepton doublets and singlets, of all generations.

On stalks $(\mathcal{O} \otimes \mathcal{A})_p$, of the sheaf, these two operator types satisfy the (symmetric or anti-symmetric ) quantum commutation relations:
\begin{align*}
[z^i_p, Z^j_p]\phi_p &= \delta^{ij} \phi_p   &  [z^i_p, z^j_p]\phi_p &= 0 = [Z^i_p, Z^j_p]\phi_p \\
\{q^k_mp, Q^l_np\}\phi_p &= \delta^{kl}\delta^{mn} \phi_p  &  \{q^k_p, q^l_p\}\phi_p &= 0 = \{Q^k_mp, Q^l_np\}\phi_p
\end{align*}
where, as usual, \{ , \} indicates the anti-commutor. Also, on open sets containing both points, $p \ne r$, all the operators either commute or anti-commute with each other. 

(Since our product creation/annihilation operators tensor a symmetric operator with an anti-symmetric operator, their overall behavior is anti-symmetric, so they obey Fermi-Dirac statistics. By contrast, the $SU(3), Sp(1), U(1)$ connection forms only operate on one of the sheaves, either $\mathcal{O}_{\ML}$, for $SU(3)$ and $U(1)_{Y_C}$, or $\mathcal{A}_{\ML}$, for $Sp(1)$ and $U(1)_{Y_S}$. Thus, since an operator is the product of a vector space and its dual, the connection forms are symmetric \xc symmetric or anti-symmetric \xc anti-symmetric, and the product is alway symmetric. So connection forms (interactions) obey Bose-Einstein statistics.)

Now we introduce our maps $\PL$: they associate points, $x$, in the space-time manifold, $\HP^1_o$, and vectors at each point, $e^0_x$, with points and vectors on \ML. 
\[
\PL : (x, e^0_x) \mapsto (\PL \xc (\PL)_n(x), \Zo \xc \Qo)_p(x))
\]

The image points have coordinates $\psi_\lambda(x)^i \otimes \psi_\lambda(x)^j_n$; the image vectors have components $(\Zo)^i \otimes (\Qo)^j$. Clearly $\psi_\lambda(x)^i \otimes \psi_\lambda(x)^j_n$ and $(\Zo)^i \otimes (\Qo)^j$ operate on the stalk $\mathcal{O}_{\psi(x)} \otimes \mathcal{A}_{\psi(x)}$. We will call $\psi_\lambda(x)^i \otimes \psi_\lambda(x)^j_n$ and $(\Zo)^i \otimes (\Qo)^j$, in their capacity as operators on $\mathcal{O}_{\psi(x)} \otimes \mathcal{A}_{\psi(x)}$, ``field operators".

\subsubsection{The Time-Evolution Operator}\label{S:prop}

The basic process in a quantum field theory experiment is that a set of states (the ``initial states'') is associated  with a position $x$ on the space-time manifold.  Then the Lagrangian density of the Standard Model is used to construct an operator on states called the time evolution operator (sometimes called the ``propagator of the interacting fields''), that sends the initial states into a different set of states, the ``final states,'' that are now associated with another location in space-time, which is causally connected to the first position. The final step requires us to have an inner product on the states at each point, so that the final states at point two can be compared with a set of test states at that point. The scalar results are called ``expectation values.'' The propagator is required to preserve the inner product. (Notice that even if the points are not causally connected, there may be non-zero expectation values. Classically, of course, non-causally connected points cannot influence each other.)

As we have defined quantum states in our sheaf, the propagator associates a curve in space-time from
$(x, e^0_x)$ to $(y, e^0_y)$ in $\HP^1_o$, with another curve in classical state space, from 
\begin{align*}
((\PL \otimes (\PL)_n)(x), (\Zo \otimes \Qo)(x)) \quad \text{to} \\
((\PL \otimes (\PL)_n)(x), (\Zo \otimes \Qo)(y)) 
\end{align*}
and then with a sheaf homomorphism 
\[
(\mathcal{O} \otimes \mathcal{A})_{\psi(x)} \to (\mathcal{O} \otimes \mathcal{A})_{\psi(y)}
\]

In order to generate causal curves that preserve the optical structure on $\HP^1_o$, we need a causal flow on $\HP^1_o$, that is parallel with respect to $\lcl{}{}$. For that we need a causal, $\lcl{}{}$-parallel vector field to integrate. The Dirac vector field, $N_g$, introduced in section \ref{S:metric}, and uniquely determined by the metric $g$, is just such a causal parallel vector field that integrates to a flow. 

Under the differential of all elements of the class $| \PL |$, $N_g$ maps to a vector field in every leaf of $E$, and thus to a vector field on $\ML$.  This vector field holomorphic and integrable on $\ML$. It therefore integrates to a holomorphic flow on $\ML$, that is, one that preserves $I$, and thus the sheaf $\mathcal{O}_{\ML}$. Thus, each Lorentzian metric $g$ on $\HP^1_o$ with a parallel spinor field is associated with a holomorphic flow on $\ML$ generated by $d \PL(N_g)|, \  \forall \PL \in | \PL |$. This holomorphic flow is an element of the set of all automorphisms of \ML. 

We know that these automorphisms can be regularly represented on the sheaf sections $\mathcal{O}(U)$ over open sets $U$ containing both $\psi_\lambda(x)$ and $\psi_\lambda(y)$. But such global sections may be hard to find. This is where quantum field theory comes in handy. The classical hamiltonian can be used to define the flow on the distribution $E$, using sub-pseudoriemannian theory. This classical hamilton is expressed in terms of the various fields we have defined.  When we replace the fields with field operators on the sheaf, as we have defined them, we automatically get a representation of the flow as a sheaf homomorphism, without reference to global sections.  This representation is the quantum field time evolution operator or propagator.

The classical propagator operates within one leaf of the distribution $E$. Let us call that one ``classical parallel universe''. However, since the sheaf is defined on open sets of \ML, germs of the sheaf, and hence quantum states touch an infinite number of parallel classical universes.  Thus, the quantum field propagator, which sends germs to germs will affect more than just one leaf of a distribution. This is why in quantum field theory there can be non-causal influences, for example.

\subsubsection{Expectation Values and Comments on QFT}

The final step of a quantum field theory experiment is to calculate expectation values, that is inner products between final states and some test states, perhaps the vacuum state. These expectation values are the actual predictions of the theory. This step requires us to define an inner product on states. Previous efforts to geometrize quantum field theory have found this step difficult, because it usually requires the use of global sections and some kind of $L^2$ metric, based on an $\ML$ measure. Such a metric can indeed be defined, but in this theory it is unnecessary, since we only use germs for states. As a result, we can make do with an $l^2$ inner product on the stalks, which is easy for germs of holomorphic functions.

As a final comment on quantum field theory: our examination has shown that the algebraic geometry of \ML, based on the structure sheaf $\mathcal{O}_{\ML} \otimes \mathcal{A}_{\ML}$ is plausibly the quantum field theory equivalent of the differential geometry of $\ML$ embodied in our classical field equations. Specifically, the different choices of $g$ on the differential side define automorphisms of $\ML$ which, on the algebraic side, are represented by homomorphisms of the structure sheaf .

\section{Conclusions and Subjects for Future Work}\label{S:fut}

The very strong hypothesis that we asserted in section \ref{S:hyp}, seems to be remarkably well supported. $\ML$ seems to define a ÒmultiverseÓ, with each choice of Brinkmann metric $g$, and associated parallel spinors defining a new class of closely related maps $|\PL|$, that together define an equivalence class of parallel classical ÒuniversesÓ with the same physical parameters.  A different $g$ will bring a different class of parallel universes with different physical parameters. Presumably, each value of $\lambda$ defines a different multiverse, with $\lambda$ affecting some physical parameter, perhaps the minimal energy values for the Higgs and coupling fields. The quantum field theory seems to be defining representations of \ML-automorphisms on a specific structure sheaf for \ML, which is therefore defined over an entire equivalence class of parallel classical universes.
  
\subsection{Subjects for Further Research}

Despite the progress, much work remains to be done.  Here are just a few short descriptions of important issues for future research:
\begin{itemize}
\item	Does our particular pp-wave Einstein equation support a non-zero minimum energy solution for the gravitational connection?

\item	Does our conclusion that $G = e^{-1}$ conflict with observations of G?

\item	Can our analysis of the Higgs field, mass matrix, and CKM, and PMNS matrices as fields be shown to be equivalent to the very elaborate Standard Model Higgs sector?

\item    Can this model be shown to be renormalizable?
\end{itemize}

\subsection{Final Speculations}

What would it mean if the hypothesis were true? 

There is a very long-lasting intellectual tradition in Western philosophy that holds that pure geometry underlies nature, that mathematics is more than just a useful tool for understanding the world, that, somehow, the universe is mathematical. Kepler and Copernicus and Ptolemy, despite their differences, agreed that planetary orbits must be conic sections of some sort, not just irregular paths. They represented a tradition in physics, going back through Plato, to Parmenides, and ultimately to Pythagoras, that we can learn about about the universe itself just by examining geometric structures through contemplation.
      
In the last few hundred years we have gone very far in another, ÒCopernicanÓ direction. We have learned that the earthÕs orbit is not a perfect conic section; and that neither the earth, nor the sun, nor the galaxy, have mathematically interesting locations in the universe.  This tradition holds that we humans are the important things in the universe, and that many of the universe's properties exist simply because without them we would not exist to observe them. This ``anthropic'' perspective sees the Pythagorean alternative as mystical hogwash, and believe that the only way we can get reliable information about a universe with many, possibly infinitely many, arbitrary parameters is to experiment and observe.
      
If the hypothesis we have advanced is true, both traditions will derive some support. On the Copernican side, the mass matrices and other constants will specify the orientation of $E$ relative to $S_n$ and $D_n$. So some, maybe even most, of the properties of our particular observed universe will be defined by the selection of the original Brinkmann metric $g$. 
      
But for the Pythagorean side, we note that the multiverse itself, \ML, the ``quaternionic Picard variety of the quaternionic elliptic curve'', seems to be very central to the geometry of four dimensions, which as we now know to be ``a dimension like no other'' \cite{as05}. And while the selection of an $E$ orientation, and thus a particular universe, is arbitrary, the overall geometry of $\ML$ is very specific, and we can learn a lot about our particular universe simply by studying \ML.

\end{document}